\documentclass{article}
\usepackage[left=2.25cm,top=2cm,right=2.25cm,bottom=2cm]{geometry}
\usepackage{amsmath,amssymb}
\usepackage[english]{babel}
\usepackage[utf8]{inputenc}
\usepackage{dirtytalk}
\usepackage{float}
\usepackage{graphicx}
\usepackage{caption}
\usepackage{subcaption}
\usepackage{hyperref}
\usepackage{amsthm}
\usepackage{multirow}
\usepackage[marginclue,footnote,silent,draft]{fixme}

\theoremstyle{definition}

\addto\extrasenglish{

}

\usepackage{natbib}

\usepackage{tikz}
\usetikzlibrary{positioning}

\title{Graph-Based Analysis and Visualisation of Mobility Data}

\author{Rafael Martínez Márquez, Giuseppe Patan\`e\\ CNR-IMATI, Italy}
\date{}
\begin{document}
\maketitle
\begin{abstract}
Urban mobility forecast and analysis can be addressed through grid-based and graph-based models. However, graph-based representations have the advantage of more realistically depicting the mobility networks and being more robust since they allow the implementation of Graph Theory machinery, enhancing the analysis and visualisation of mobility flows. We define two types of mobility graphs: Region Adjacency graphs and Origin-Destination graphs. Several node centrality metrics of graphs are applied to identify the most relevant nodes of the network in terms of graph connectivity. Additionally, the Perron vector associated with a strongly connected graph is applied to define a circulation function on the mobility graph. Such node values are visualised in the geographically embedded graphs, showing clustering patterns within the network. Since mobility graphs can be directed or undirected, we define several Graph Laplacian for both cases and show that these matrices and their spectral properties provide insightful information for network analysis. The computation of node centrality metrics and Perron-induced circulation functions for three different geographical regions demonstrate that basic elements from Graph Theory applied to mobility networks can lead to structure analysis for graphs of different connectivity, size, and orientation properties.
\end{abstract}


\tableofcontents

\section{Introduction\label{section:introduction}}
The mobility of individuals is a topic of critical importance for the development and sustainability of cities. Forecasting the flows of crowds in a city and foreseeing possible drawbacks in the transport system can boost public decision-making in risk mitigation and optimising traffic flow, thus leading to public safety and citizens' well-being. Traffic flow is affected by various factors, such as the spatial dependencies between the different regions of a city, the temporal dependencies from near or distant past time intervals, and the effect of external factors, such as weather, land and road features.

The movement of people and vehicles within a city and region is increasingly influenced by interdependence. In particular, the optimal use of different transport infrastructures and networks in cities and their connection to regional destinations and between neighbouring regions has a substantial impact on various aspects, such as chemical and noise pollution, the inefficiency of local public transport due to a general increase in traffic volume, the risk of accidents, and high emergency management costs. Concerning state-of-the-art in traffic and mobility data-driven modelling, we identify two main approaches (\autoref{section:models}): Euclidean models that use grid partitions over a region for forecasting the flows within them~\citep{stresnet_zhang,finegrids_liang} or to classify the available data into training and testing sets~\citep{deepgravity_pappalardo}, and Non-Euclidean models that use graph-based structures~\citep{zikaizhang_multipledynamic_speed,shaokun_gat_flow,taoliu_dynamicmodelling_urbanroadnetwork} that would be more suitable and realistic for mobility modelling as the spatial dependency is not necessarily between adjacent cells in a grid-based partition. 

For both Euclidean and non-Euclidean models, visualising the predicted flows becomes an essential tool for traffic analysis because it provides additional interpretations to the purely theoretical ones. For instance, visualised traffic flows can help to localise and identify traffic patterns or areas with the highest or lowest traffic values (e.g. speed, volume). In addition to the visualisation of mobility flows and volumes, the node centrality metrics (i.e., Closeness centrality, betweenness centrality, and Page rank centrality) (\autoref{sec:GRAPH-OVERVIEW}) provide further analysis tools for the structure of the system, particularly within a graph framework. For instance, the betweenness centrality can be a measure of the vulnerability of a particular station to traffic disruption~\citep{metro_scalzo, roadindia_mukherjee}; the closeness centrality can be used to find the stations that are best connected to the urban network in terms of minimising the total distance, travelling time, or other road parameters. The Page rank centrality can identify the most popular roads or stations regarding the dynamic mobility flow within the urban network. By analysing both the traffic flow and the structure of the network, it is also possible to better understand how the network structure affects the dynamic of the traffic; other factors are operational interventions and seasonal factors that affect the travel frequency on the network. We introduce node centrality metrics analysis for two types of graphs: Region Adjacency graphs and Origin-Destination graphs. The former type of graph is induced by a given finite partition of a city or region, where the connections represent the adjacency of a pair of partition units. The latter type of graph is induced by an Origin-Destination matrix, a square matrix (possibly sparse) whose entries are positive values representing a mobility counting between a pair of sub-regions (not necessarily geographically adjacent). We compute, visualise, and analyse the three centrality metrics for the Region Adjacency graphs of Genova Province, United Kingdom and New York State. Additionally, we perform the same visual analysis for the Origin-Destination graph for the flows predicted by the Deep Gravity mobility model~\citep{deepgravity_pappalardo}.

Furthermore, through some matrices associated with a graph, e.g. the Laplacian matrix (\autoref{sec:GRAPH-MATRICES}), we can investigate demographic factors that affect mobility flows such as the distribution of population surrounding each node of a transport network~\citep{metro_scalzo}.  Also, the Laplacian spectrum is fundamental for Graph Signal Processing tools that facilitate the analysis of the correlation of time series of traffic volumes, thus identifying data correlation structures over different locations and Origin-Destination pairs in an urban network. The classical graph Laplacian is the Combinatorial Laplacian, which satisfies important properties as symmetric and positive semi-definite. However, these features hold only for undirected graphs because they are based on the symmetry of the adjacency matrix. We introduce the definitions of the  Combinatorial Laplacian and Normalised Laplacian for undirected graphs. Similarly, the Combinatorial Directed Laplacian, Symmetrized Laplacian, Combinatorial Symmetrized Laplacian, and Diplacian are defined for directed graphs. The Symmetrized Laplacian, Combinatorial Symmetrized Laplacian, and Diplacian are defined only for strongly connected graphs since they are based on the Perron vector associated with the transition probability matrix, and their existence and uniqueness are guaranteed when the graph has only one connected component. These definitions for the Laplacian matrix grants their characteristic properties to hold regardless of the graph's connectivity or if it is directed or undirected.

Overall goals we aim to present a variety of data-driven traffic and mobility models and develop a set of methods for analysing and visualising information about the dynamics of mobility flows. The input data used for such models is mainly consisting of trajectories of cars, bicycles, or people~\citep{stresnet_zhang,finegrids_liang}, sensors and monitoring stations~\citep{zikaizhang_multipledynamic_speed,shaokun_gat_flow,taoliu_dynamicmodelling_urbanroadnetwork}, and demographical data~\citep{deepgravity_pappalardo}.  The primary research and development activities include

\begin{itemize}
	\item an overview of Grid-based and Graph-based data-driven traffic models;
	\item the visualisation of centrality metric values in mobility graphs to display an overview of the network structure to more easily identify mobility flows of interest;
	\item the definition and properties of distinct Laplacians matrices for undirected and directed graphs.
\end{itemize}

These tools are essential to support autonomous decision-making for traffic management in areas of interest, e.g., to propose suggested routes to commuters entering a city to reduce traffic in certain areas based on traffic forecasts. We draw several conclusions, indicate possible extents of this work (\autoref{sec:FUTURE-WORK}), and describe the used data sets.

\section{Previous work on data-driven traffic models\label{section:models}}
The traffic forecasting problem can be formulated as follows: given a historical collection~$\mathcal{X}$ of features, e.g., average speed, road volume, and demand levels, during~$P$ different time intervals, predict a collection~$\mathcal{Y}$ of features (possibly different) for the next~$Q$ time intervals. This relation is expressed by \mbox{$\mathcal{Y}=f(\mathcal{X})$}, where~$f$ is an unknown function. Several data-driven traffic models depend on the available data and the predicted variable. Also, these models can be distinguished based on how they structure the data (e.g., in a regular or irregular domain). Some grid-based approaches are presented where the data is partitioned into squared cells, and the predicted value is the number of flows within their inner sub-partitions. Similarly, we introduce some graph-based models where data partition depends on the distribution of the road network or data collectors (e.g., sensors, monitoring stations, road junctions). These graph-based models can predict various traffic states (e.g., speed, flow, occupancy). If the chosen model is grid-based, the city or area of interest is partitioned into an~$I\times J$ grid with~$N=IJ$ cells. If the chosen model is graph-based, then the road network over a city or area of interest is represented by~$N$ nodes and a set of edges connecting them. For both grid-based and graph-based models, different techniques from Deep Learning can be used, such as convolution networks and residual units~\citep{stresnet_zhang, finegrids_liang}, or graph attention mechanisms and spectral convolution networks~\citep{gacan_graph_zhang,multistep_graph_convolution}. 

\subsection{Grid-based traffic models\label{sec:GRID-MODELS}}
Several flow forecasting models use different types of mobility data and prediction techniques. Euclidean approaches for flow prediction in traffic modelling consist of partitioning a city, state, or country into a square or rectangular grid to assess the number of displacements between two elements of the grid~\citep{stresnet_zhang,finegrids_liang}, or dividing available geographical data into training, and testing data for the use of a Deep Learning model~\citep{deepgravity_pappalardo}. When partitioning the city into an~$I\times J$ grid, urban flows, traffic volumes, and other mobility values, data can be stored in matrices in~$\mathbb{R}^{I\times J}$, or tensors in~$\mathbb{R}^{L\times I\times J}$ if the data consists of more than one type of value, where~$L$ is the number of different values that compose the data. The forecasting problem can be stated as given historical observations of mobility data~$\{\mathbf{X}_1,\mathbf{X}_2,\hdots,\mathbf{X}_N\}$ during~$N$ time intervals, predict~$\mathbf{X}_{N+1}$. External factors may vary over time and influence mobility in a city. Indeed, the crowd flows can be represented by tensors in~$\mathbb{R}^{K\times I\times J}$, where~$K$ is the number of external influences being considered. Suppose the~$I\times J$ grid is only used to classify the available mobility data. In that case, each of the cells should be, in addition, sub-partitioned (possibly in irregular units) to forecast the flows from one to another within the same cell.

\subsubsection{ST-ResNet - Spatio-Temporal Residual Network\label{sec:SPATIO-TEMP-MODEL}}
In the \textit{ST-ResNet} model (\textit{Spatio-temporal residual network})~\citep{stresnet_zhang}, the subdivision of the city into regions of interest is performed by a squared grid partition using the longitudes and latitudes. The predicted traffic variables are the inflows and outflows among the cells of the squared grid. \textit{Inflow} is the total traffic crowds entering a cell in the grid from any other cell during a given time interval. \textit{Outflow} denotes the total traffic of crowds leaving a cell to any other cell in the grid during a given time interval. The input data for the ST-ResNet model consists of trajectories of cars, bicycles, people, etc., during different constant length time intervals and meteorological data as external factors that impact the traffic flow in a city. Formally, the city, state, or country is split into an~$I\times J$ mesh with~$N=IJ$ cells, and we can associate the inflow and outflow of crowds to every cell~$(i,j)$ at the time interval~$t$. 

The ST-ResNet model applies convolution-based residual networks for dealing with the local and global spatial dependencies that affect the flows within a city. There are three such networks, one for each type of temporal dependency: closeness, period, and trend. Moreover, the ST-ResNet model uses a fully connected network to include the effect of external factors that may impact traffic flows, such as weather conditions and holidays. All these model elements are dynamically aggregated to produce predicted inflow and outflow matrices with the exact dimensions of the squared grid partition of the city. 

\subsubsection{STRN - Spatio-Temporal Relation Network\label{sec:SPATIO-TEMP-RELATION}}
The STRN model (\textit{Spatio-Temporal Relation Network})~\citep{finegrids_liang} is another grid-based model for flow prediction. The subdivision of the city into a squared grid partition consists of a considerably larger number of cells~$N$ (of smaller size) than with the ST-ResNet model; this type of partition is called \textit{fine-grained} or \textit{high resolution}. Moreover, a secondary irregular partition of the city into~$M$ regions, with~$M$ considerably smaller than~$N$, is used to assess the effects of global spatial dependencies more efficiently, namely, by analysing the diffusion within the connectivity network that represents the irregular partition, which is computationally faster than using a sequence of convolution layers as in the ST-ResNet model. The input data for the STRN model consists of trajectories as in the ST-ResNet model. 

An assignment matrix~$\mathbf{B}\in \mathbb{R}^{N\times M}$ is used to connect the~$N$ cells grid structure with the~$M$ region's network, where the entry~$b_{ij}$ denotes the likelihood that the cell~$i$ belongs to the region~$j$. The irregular partition can be computationally generated by additional available information, such as the road network, the administrative divisions of a city, or census areas. The STRN model applies a \textit{Meta Learner} that converts the features related to external factors to a representation with the same dimensionality as the \textit{Inflows} and \textit{Outflows} from the three temporal dependencies, e.g., closeness, period, and trend. Then, these four feature representations are passed to a \textit{Backbone Network} consisting of squeeze-and-excitation networks, and convolution layers, whose output is finally passed to the \textit{Global Relation Module (GloNet)}, which converts the grid-based representations to the network structure and applies \textit{Graph Convolution Networks} to assess the diffusion of the flows within the irregular regions. Then, the Meta Learner converts the predictions to the original grid-based structure. 

\subsubsection{Deep Gravity Mobility Model\label{section:deepgravity}}
The \textit{Deep Gravity mobility model}~\citep{deepgravity_pappalardo} is a grid-based model for flow prediction without using the three-time dependencies paradigm (i.e., closeness, period, trend), which is fundamental in the ST-ResNet and STRN models. To use the Deep Gravity model, partitioning a city, state, or country starts with constructing a square grid covering the whole region. Additionally to this handmade regular partition, it must be available a secondary finer (and irregular) partition for which there is available population data, for instance, the tracts or areas that constitute the units of a census. Then, the barycentre of each irregular unit is allocated in one of the square cells. Hence, the data of a cell consists of the data of the irregular units whose barycentre belongs to that cell. The square cells are divided into training and testing cells in a stratified manner based on the population data in the cells so that the two groups have the same number of cells belonging to the various population deciles. Newton's law of universal gravitation inspires this method, which is why its name is inspired. A generalised version of this attraction law, by considering the \textit{deterrence function}, is defined as

\begin{equation}
\label{eq_singly_dg}
y(l_i,l_j)=O_i\dfrac{m_j^{\beta_1}f(r_{ij})}{\sum_k m_k^{\beta_1}f(r_{ik})},   
\end{equation}

where~$y(l_i,l_j)$ represents the predicted flow from a location~$l_i$ to a location~$l_j$, and~$r_{ij}$ is the distance between them; the~$m_i$ are the populations, and~$\beta_1$ is a real parameter. The relation in Eq.~\eqref{eq_singly_dg} is called a \textit{singly-constrained gravity model} since it requires knowing the total outflow~$O_i$ for each location in advance. The Deep Gravity model applies \textit{feed-forward neural networks} to predict the number of flows from a given area from the irregular partition to any other of the areas within the same cell. The choice of this type of network lies in the fact that feed-forward neural networks generalise linear models as in Eq.~\eqref{eq_singly_dg}. Unlike the ST-ResNet and STRN models, the Deep Gravity model does not use the grid structure to predict flows between them but to split the city into training and testing regions.

The city's subdivision into a squared grid partition splits the available data sets into training and testing data since the predicted flows correspond to the flows between additional irregular partitions within each cell provided by census areas. The input data for the Deep Gravity model consists of population values obtained from the official census and geographic features obtained from \textit{OpenStreetMap}. The intuition of this model is that the flow between two locations is directly proportional to their population and inversely proportional to the distance between them.

Given two irregular units~$l_i$ and~$l_j$ in a cell~$C$, the components of the input vector~$x_{ij}$ used to predict the flow~$y(l_i,l_j)$ from~$l_i$ to~$l_j$ consist of the population in~$l_i$ and~$l_j$, the distance between them, and eighteen geographical values extracted from OpenStreetMap (land use areas, road lengths, counting of points of interest, etc.). The input vector~$x_{ij}$ passes through a multilayer feed-forward neural network with an output layer of dimension~$1$ (a scalar~$y_{ij}$) in the range~$(-\infty,\infty)$. After computing~$y_{ij}$ for~$j=1,\hdots,M$, where~$M$ is the number of irregular units in the cell~$C$, the~$M$ values are passed through a \textit{softmax} layer to convert them into an~$M$-dimensional vector with non-negative components whose sum is~$1$, namely, a probability vector that represents the probability distribution of the flows starting in location~$l_i$ within the cell~$C$. When training the model, the predicted values~$y_{ij}$ are compared to real flows~$z_{ij}$ from location~$l_i$ to location~$l_j$, for instance, provided by GPS trajectory data or commuting surveys. A standard metric used to measure the performance of a flow prediction model is the \textit{Common part of Commuters (CPC)} defined by

\begin{equation*}
CPC=\dfrac{2\displaystyle \sum_{i,j}\min(y_{ij},z_{ij})}{\displaystyle \sum_{i,j}y_{ij}+\sum_{i,j}z_{ij}},
\end{equation*}

where~$y_{ij},z_{ij}$ refer to the predicted and real flow from location~$l_i$ to location~$l_j$ respectively, and the indices~$i,j$ run along all the locations in a region of interest, for instance in one of the square cells used in the Deep Gravity model. The CPC values are in the range~$\left[0,1\right]$ with~$1$ indicating a perfect flow prediction. Suppose the total number of outflows in a region of interest coincides with the predicted flows (as with the Deep Gravity model used by~\citep{deepgravity_pappalardo}). In that case, the CPC value is equal to the fraction of the flows correctly predicted by the model. 

\subsection{Graph-based traffic models\label{sec:GRAPH-MODEL}}
Graph-based approaches have been used for metro traffic modelling~\citep{metro_scalzo} and statistical analysis of road networks~\citep{roadindia_mukherjee}. Graph theory provides additional theoretical elements, namely the weight of the edges, which can be used to represent some features in traffic models, such as the road length and road capacity~\citep{complexnetwork_tian}. Moreover, the centrality properties of a graph can provide additional insight into the distribution and importance of nodes and edges~\citep{centralitymetrics_henry,combinatorial_boulmakoul}.

There are several approaches to creating a graph structure for traffic modelling depending on the available data type and the forecasted variable. In a traffic graph~\citep{jiang_graph_neural_survey}, a \textit{graph signal}~$\mathbf{X}_t\in\mathbb{R}^{N\times d}$ is defined at every time step~$t$ where~$N$ is the number of nodes and~$d$ is the number of \textit{traffic elements} (e.g. speed, traffic state, traffic demand) that are measured. The majority of graph-based spatio-temporal traffic problems belong to traffic state and traffic demand prediction~\citep{survey_ieee}, which are modelled by various graph-based deep learning architectures (e.g. \textit{Spectral Graph Convolution (SGCN), Gated Recurrent Unit (GRU), Graph Attention Network (GTA)}). The nodes of a traffic graph can be defined as the intersections between roads or as the sensors along a highway. The physical road connections between them usually give the edges, and the predicted variables can be any traffic element for each node, namely, a graph signal.

\subsubsection{ETGCN - Evolution Temporal Graph Convolutional Network\label{section:etgcn_graph_speed}}
The \textit{Evolution Temporal Graph Convolutional Network (ETGCN)}~\citep{zikaizhang_multipledynamic_speed} captures the spatial and temporal correlations among the nodes in a \textit{Road graph}. The road graph is a weighted graph whose nodes are traffic sensors distributed over a city or region. The concept of adjacency matrix involves the fusion of three types of information (Content Similarity Adjacency Matrix, Graph Betweenness Adjacency Matrix, and Transportation Neighborhood Adjacency Matrix) whose entries consider the geographical position of the sensors and their connections in the original road network, and which fusion enhances feature learning. Given a historical time series of the registered speed at every node, the ETGCN model aims to predict the speed at every node in the next timestep. The architecture of the ETGCN model combines Graph Convolutional Networks and GRU to learn the sequence of spatial and temporal features.

\subsubsection{GTA - Graph-based Temporal Attention Framework\label{section:gta_graph_flow}}
The \textit{Graph-based Temporal Attention Framework (GTA)}~\citep{shaokun_gat_flow} model also considers a weighted sensor graph, but in contrast to the ETGCN model, the weights of the edges are the road network distances between the sensors, and the predicted variable is the \textit{traffic flow}, which is defined as the number of vehicles passing through the monitoring station over a given time interval. The architecture of the GTA model introduces an attention mechanism to adaptively identify the relations among three temporal dependencies: monthly pattern, weekly pattern, and current pattern. A \textit{Long Short-Term Memory (LSTM)} network is employed to extract the temporal correlation for each dependency.

\subsubsection{G-STARIMA - Graph-based Spatio-Temporal ARIMA Model\label{section:gstarima_graph_flow}}
The \textit{Graph-based spatio-temporal ARIMA (G-STARIMA)} model~\citep{taoliu_dynamicmodelling_urbanroadnetwork} is a graph-based framework built upon statistical methods. The road network is described by a weighted undirected graph where the nodes are the traffic intersections, and the edges are their road connections. The G-STARIMA model uses a historical time series of graph signals at the network nodes (i.e., sensor observations such as speed, occupancy, or traffic flow) to predict the traffic state at the next instant. Since the connectivity between traffic intersections, represented by edge weights, may vary over time because of the dynamics of urban traffic states, it is performed a dynamic estimation of a weighted adjacency matrix which is found by solving a convex optimisation problem. 

\subsection{Discussion and comparison\label{sec:GRAPH-DISCUSSION}}
The ST-ResNet, STRN, and Deep Gravity models are grid-based models for flow predictions within the units of a city, state, or country partition. The ST-Resnet and the STRN models consider the \textit{CPT Paradigm} (closeness, period, and trend time dependencies) and predict the flow values as the next value of a sequence of flows. In contrast, the Deep Gravity model only predicts a value for the flows from a set of given features. Regarding the network architectures that build the models, the Deep Gravity model has a simpler architecture since feed-forward neural networks are the elementary Deep Learning network. In contrast, the convolution layers and residual units of the ST-ResNet and STRN models are more refined networks. Moreover, even if the three models are based on a grid structure, they use the square cells differently: the ST-ResNet and the STRN models use the cells as the units where the flows will be predicted from historical values of the same flows. In contrast, the Deep Gravity model uses the cells only to split the available data into training and testing data. 

Regarding the spatial dependencies, the Deep Gravity model performs flow prediction only for irregular units within the same cell. Therefore, it does not consider global spatial dependencies as the ST-ResNet and STRN models. The three show significant improvements concerning other existing baseline models for flow prediction. Still, it is necessary to make a performance comparison between the Deep Gravity model and the other two grid-based models. 

The grid-based models show interesting results in traffic forecasting. However, some features could suggest modelling mobility in a city as a graph: the structure of the roads as graph topology that connects different locations, i.e., graph nodes, in the city, or the information associated with movement between two locations as graph weights. The urban road network is a typical spatial network because of its geographical factors~\citep{analysis_urban_tian, survey_ieee}, and usually, adjacent cells of the grid do not have flows between them during some intervals of time because there are no direct roads connecting them. 

The ETGCN and GTA models are data-driven graph-based models for traffic predictions that consider sensors or monitoring stations as the nodes of a weighted graph. The GTA model uses the distance between the sensors as edge weights and predicts the number of vehicles passing through every monitoring station. In contrast, the ETGCN model considers a fusion of different types of adjacency matrices that enhance the learning of features and predict the speed at every network node. Another characteristic of the GTA model is the implementation of three-time dependencies, which is analogous to the CPT Paradigm of the ST-ResNet and STRN grid-based models. Regarding the Deep Learning architecture of these graph-based models, the ETGCN uses GRU networks to learn temporal features. In contrast, the GTA model combines LSTM networks through an attention-based mechanism. In contrast to the ETGCN and the GTA models, the G-STARIMA model is not a Deep Learning model but a graph-based framework based on statistical methods. In addition, this model considers the road intersections as network nodes instead of the monitoring stations and predicts various traffic states (e.g., speed, traffic flow).

Both grid-based and graph-based traffic models show interesting results, and the choice of a modelling framework should be based on the type and amount of available data and the variables that will be predicted. However, graph-based models have additional theoretical tools to boost traffic analysis, i.e. the centrality metrics (\autoref{sec:CENTRALITY-METRICS}). For instance, the betweenness centrality was used in~\citep{metro_scalzo} to analyse the robustness of the metro network in London since it facilitates the identification of metro stops that might significantly impact the whole network during a disruption. Similarly, the betweenness metric was used in~\citep{roadindia_mukherjee} to identify potential congestion points in the Indian highway network. In~\cite{survey_ieee}, various approaches are proposed to construct a graph-based framework, different mobility data sets, and Deep Learning networks for graph-based models.

\section{``Generic'' and traffic graphs: definition~$\&$ metrics\label{sec:GRAPH-OVERVIEW}}
The graph-based approaches for traffic and mobility modelling certainly build upon the fundamental concepts from Graph Theory. Basic definitions such as the direction and weight of an edge enable the construction of different types of graphs. This leads to the implementation and interpretation of diverse multi-agent systems modelling, where the connection between the agents (the nodes) stores significant information. The sequences of consecutive edges motivate the definitions of paths, which induces the concept of connected components. Furthermore, all these structures combined enable the introduction of several quantitative scores for each node according to its importance in the connectivity and influence to the rest of the network, which is characterised by the centrality metrics. In traffic and mobility modelling, the centrality metrics provide valuable information for the structure analysis of the road and geographical networks.

\subsection{``Generic'' graphs\label{sec:GENERIC-GRAPH}}
To represent a graph~$\mathcal{G}$ with nodes set~$V=\{1,\hdots,N\}$ and edges set~$E$, we use the notation~$\mathcal{G}=(V,E)$. Undirected graphs represent only connections between nodes, so~$(u,v)=(v,u)$ for every~$u,v\in V$. A graph is a \textit{directed graph} (or \textit{digraph}) if the couples that define the edges in~$E$ cannot be arbitrarily ordered, namely, if in general~$(u,v)\neq (v,u)$. The positive value~$w_{uv}$ is the \textit{weight} of the edge~$(u,v)$, and~$E$ is a \textit{weighted graph}. If no values are associated with a graph's edges, it is called \textit{unweighted}. The weights of the edges of a graph with~$N$ nodes can be stored in an~$N\times N$ matrix called the adjacency matrix of~$\mathcal{G}$.

We can associate positive weights to every edge of a graph with~$N$ nodes and store them in the weighted adjacency matrix~$\mathbf{A}\in \mathbb{R}^{N\times N}$, namely

\begin{equation*}
	A_{ij}=\left\{\begin{array}{cl}
		w_{ij}&\text{if }(i,j)\in E;  \\
		0&\text{otherwise}. 
	\end{array}\right.
\end{equation*}

If~$\mathcal{G}$ is an unweighted graph; then we have binary weights, i.e.,~$1$ for existing connections between nodes, and~$0$ otherwise. For undirected graphs, since~$(u,v)$ and~$(v,u)$ refer to the same edge, then the weight values satisfy~$\omega_{uv}=\omega_{vu}$, the corresponding adjacency matrix is symmetric. Knowing the adjacency matrix of a graph, we know the weights of the edges and the connections between the nodes. Indeed, we write~$\mathcal{G}=(V,E,\mathbf{A})$ to make more explicit the relationship between the graph representation of~$\mathcal{G}$ and the matrix representation given by the adjacency matrix~$\mathbf{A}$.

\paragraph*{Connected and strongly connected graphs}
An undirected graph is \textit{connected} if, for every pair of nodes $u,v\in V$ there exists a path~$w$ from~$u$ to~$v$, otherwise~$\mathcal{G}$ is called \textit{disconnected}. A directed graph~$\mathcal{G}$ is strongly connected if for any pair \mbox{$(p_{i},p_{j})$} of~$\mathcal{G}$ there exists a path (i.e., a sequence of edges) \mbox{$(p_{i},p_{l_{1}})$}, \mbox{$(p_{l_{2}},p_{l_{3}})$},~$\ldots$, \mbox{$(p_{l_{r-1}},p_{j})$}, which leads from~$p_{i}$ to~$p_{j}$. In this case, the path has length~$r$. The associated directed graph \mbox{$\mathcal{G}(\mathbf{A})$} of a \mbox{$n\times n$} matrix~$\mathbf{A}$ consists of~$n$ nodes \mbox{$p_{1},\ldots, p_{n}$}, where an edge leads from~$p_{i}$ to~$p_{j}$ if and only if \mbox{$A_{ij}\neq 0$}. As main properties, we recall that a matrix~$\mathbf{A}$ is irreducible if and only if the associated directed graph \mbox{$\mathcal{G}(\mathbf{A})$} is strongly connected. Otherwise, the matrix is called \emph{irreducible}. A matrix \mbox{$\mathbf{A}\in\mathbb{R}^{n\times n}$} is said to be \emph{reducible} if there exists a perturbation matrix~$\mathbf{P}$ such that

\begin{equation*}
\mathbf{C}
=\mathbf{P}\mathbf{A}\mathbf{P}^{\top}
=\left[
\begin{array}{cc}
\mathbf{A}_{11} &\mathbf{A}_{12}\\
\mathbf{0}		&\mathbf{A}_{22}
\end{array},
\right]
\end{equation*}

\mbox{$\mathbf{A}_{11}\in\mathbb{R}^{n\times r}$}, \mbox{$\mathbf{A}_{12}\in\mathbb{R}^{(n-r)\times (n-r)}$}, and \mbox{$\mathbf{A}_{22}\in\mathbb{R}^{n\times (n-r)}$}. Let~$\mathbf{A}$ be an irreducible matrix. Then, (i)~$\mathbf{A}$ has a positive real eigenvalue equal to its spectral radius \mbox{$\rho(\mathbf{A})$}; (ii) \mbox{$\rho(\mathbf{A})$} is a single eigenvalue of~$\mathbf{A}$ and its corresponding eigenvector \mbox{$\phi>0$}; (iii) \mbox{$\rho(\mathbf{A})$} increases when any entry of~$\mathbf{A}$ increases; and (iv) there is no other non-negative eigenvector of~$\mathbf{A}$ different from~$\phi$. Finally, a digraph is called \textit{weakly connected} if its underlying undirected graph is connected. A disconnected graph can be decomposed into smaller subgraphs connected as an independent graph, called \textit{connected components}.

\paragraph*{Graph paths}
A \textit{path}~$w$ in a graph (or a \textit{walk})~\citep{laplacians_chung} of length~$m$ is a sequence of different~$m$ nodes in~$V$, namely,~$w=\{v_1,\hdots,v_m\}$ such that the couple~$(v_i,v_{i+1})$ belongs to the edges set~$E$ for every~$i=1,\hdots,m-1$. Given the nodes~$a$ and~$b$ of a graph (undirected or directed), then~$w=(v_1,\hdots,v_m)$ is a \textit{shortest path} from~$a$ to~$b$ if~$v_1=a,v_m=b$ and the value \mbox{$d_w(a,b):=\sum_{i=1}^{m-1} A_{v_iv_{i+1}}$} attains its minimum among all the paths starting in~$a$ and ending in~$b$. Since there may be more than one shortest path from a node~$a$ to a node~$b$, we define the distance from~$a$ to~$b$ as \mbox{$d(a,b)=\min \{d_w(a,b)\colon w \text{ is a path from }a\text{ to }b\}$}, which is a unique value. If for a couple of nodes~$a,b\in V$ of a disconnected graph, there is no path starting in~$a$ and ending in~$b$ then we define their distance as~$d(a,b)=\infty$. The distance between two nodes is not precisely a distance in the context of metric spaces when the graph is directed because~$d(a,b)$ is not necessarily the same as~$d(b,a)$.

\begin{figure}[t]
	\centering
	\begin{tabular}{ccc}
		\multicolumn{3}{c}{Undirected Region Adjacency graphs}\\
		\hline
		\multicolumn{3}{c}{\ }\\
		\includegraphics[scale=0.20]{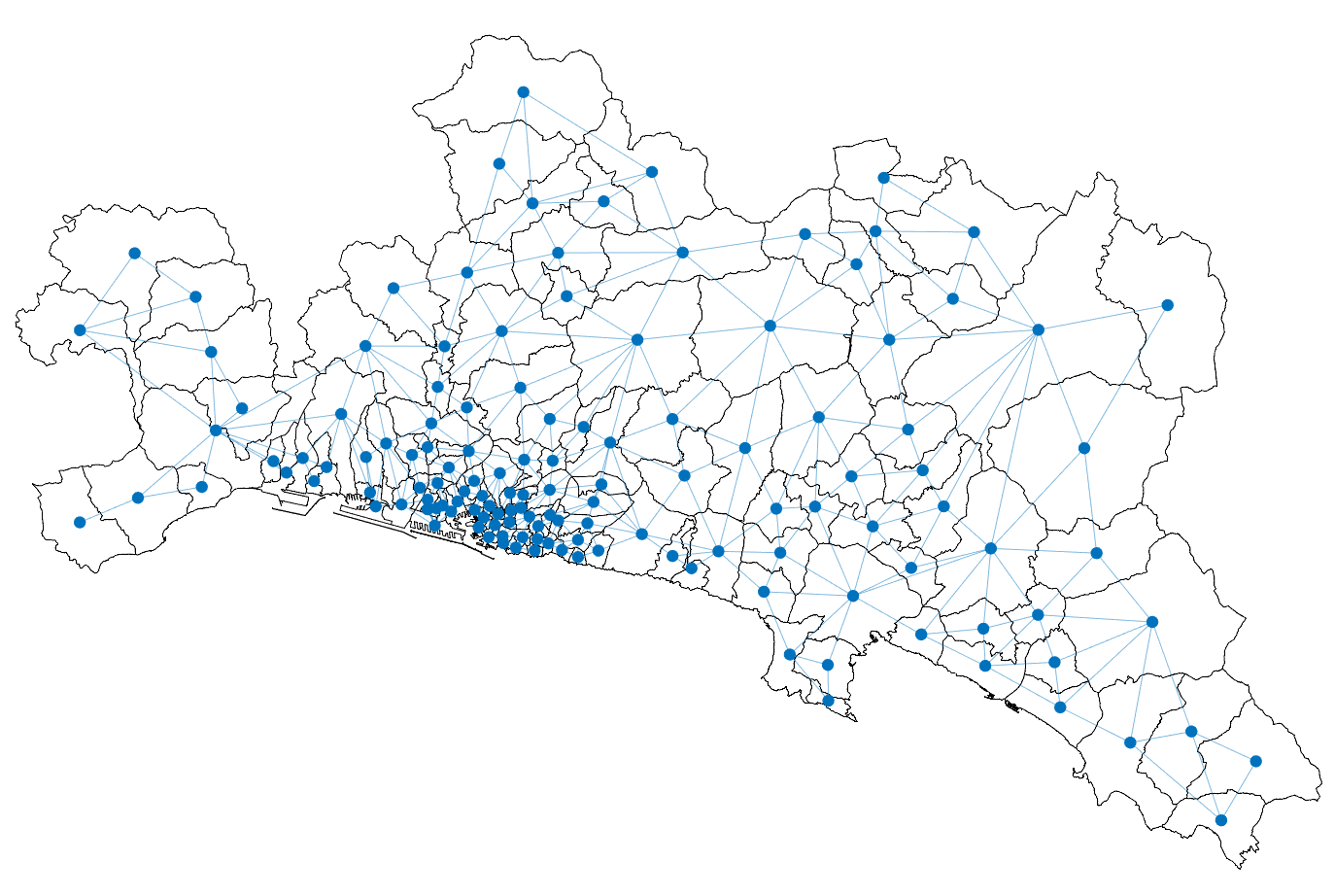}
		&\includegraphics[scale=0.20]{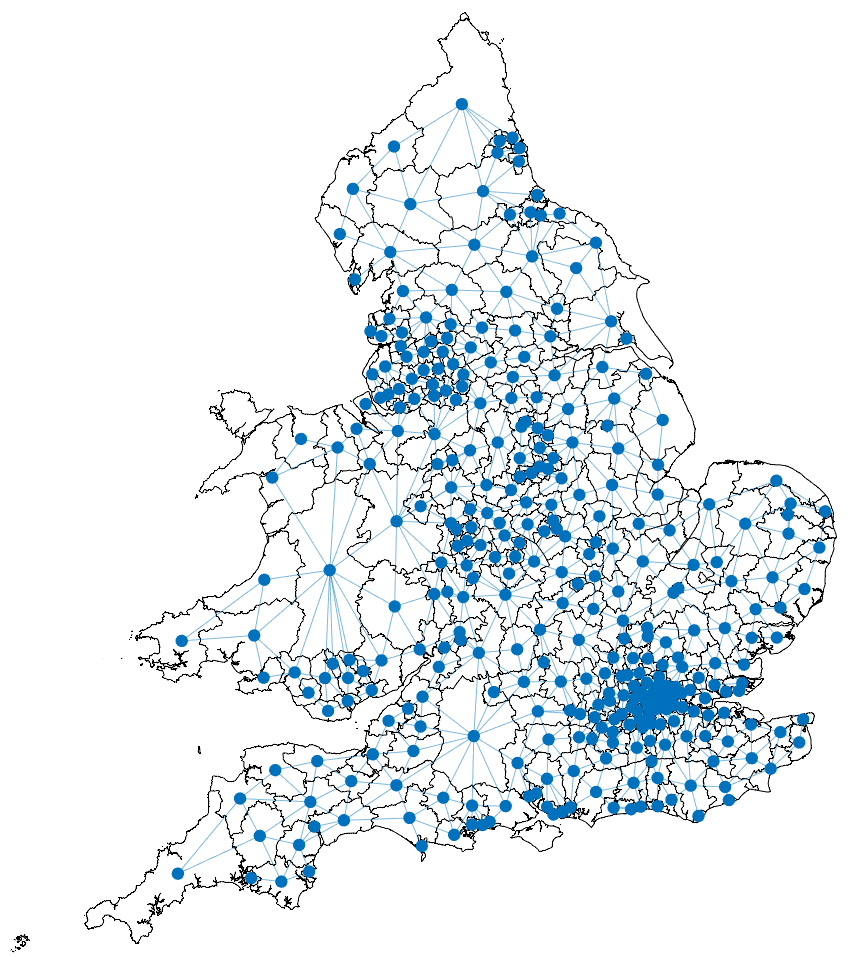}
		&\includegraphics[scale=0.20]{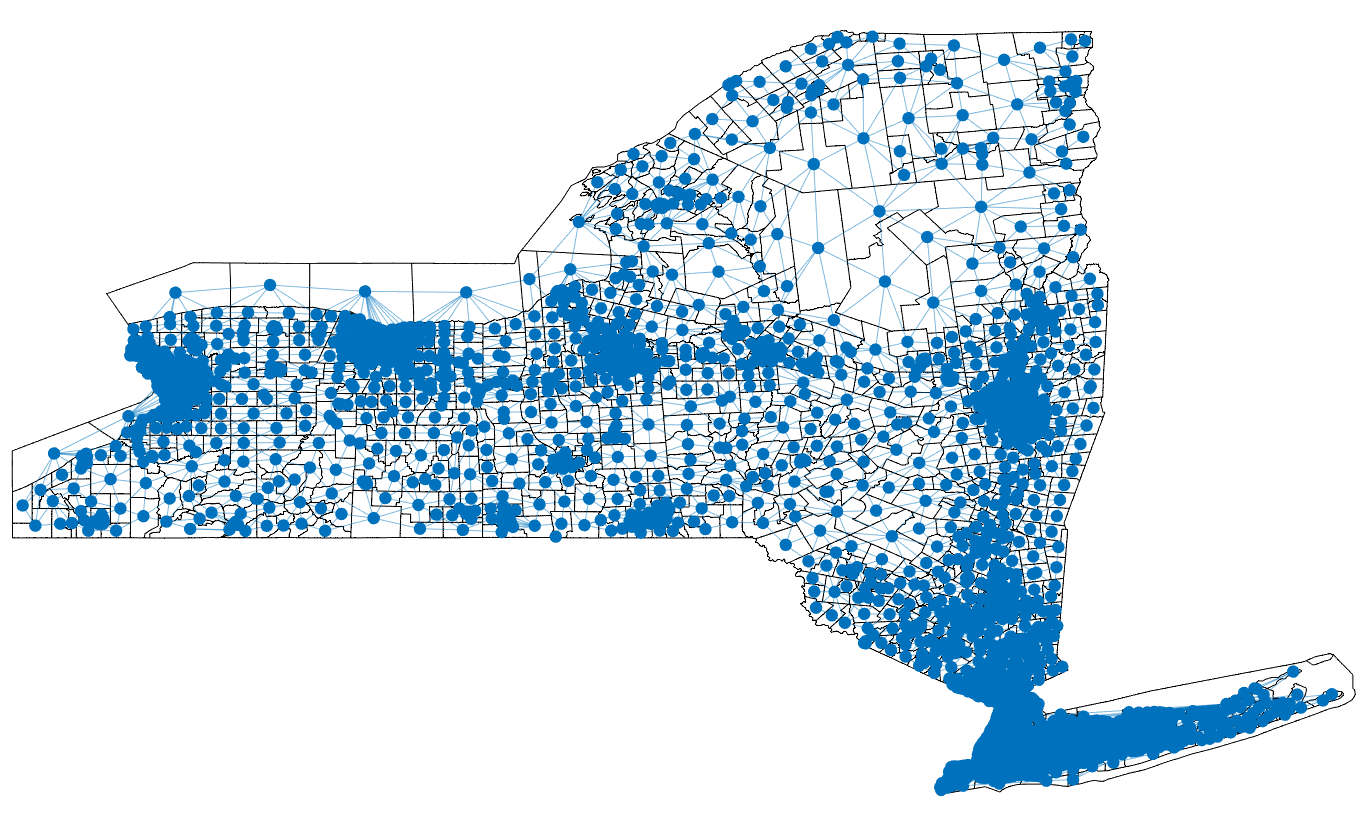}\\
		\multicolumn{3}{c}{Region Adjacency graph}\\
		\hline
		\multicolumn{3}{c}{\ }\\
		\includegraphics[scale=0.20]{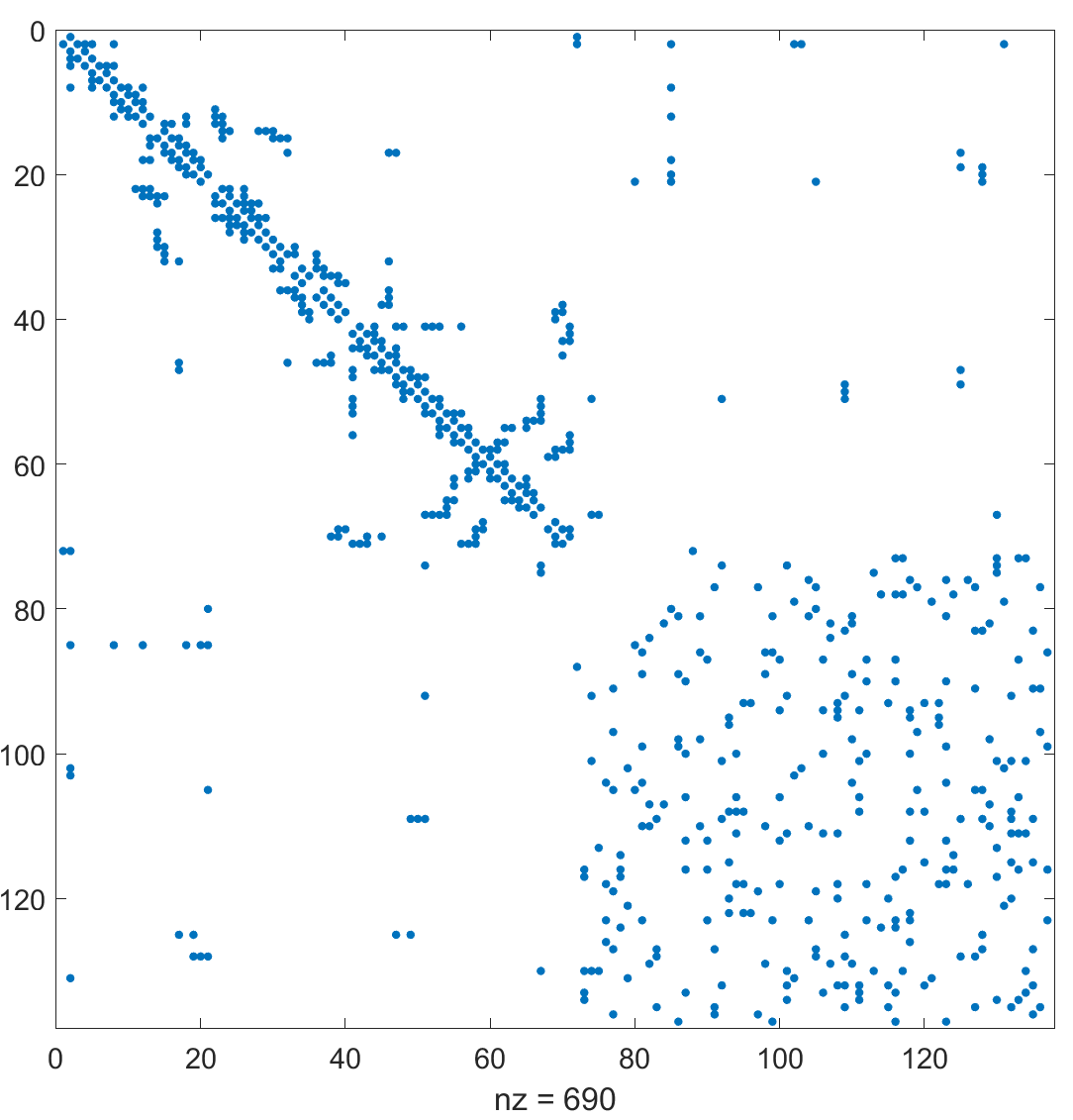}
		&\includegraphics[scale=0.20]{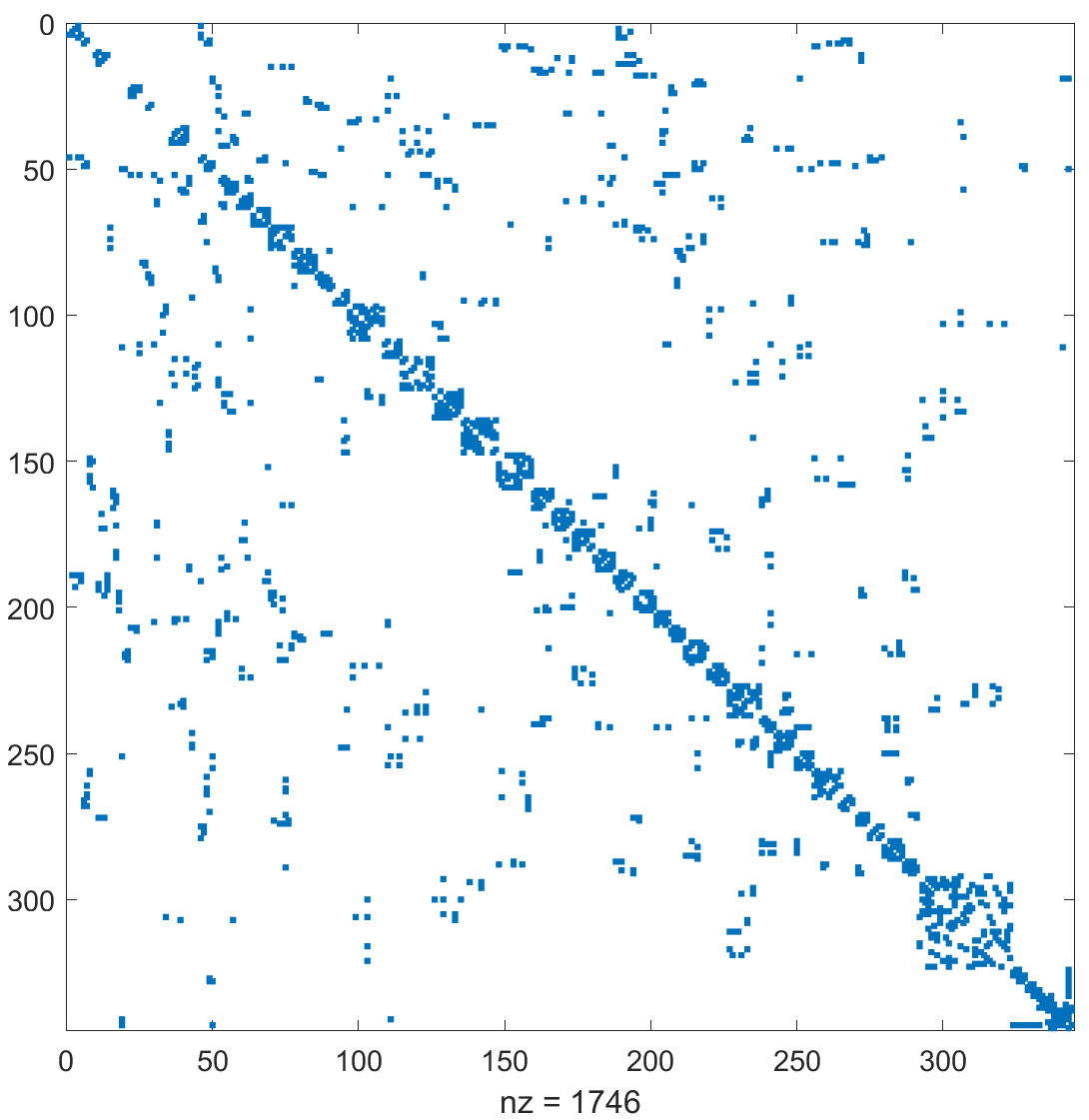}
		&\includegraphics[scale=0.20]{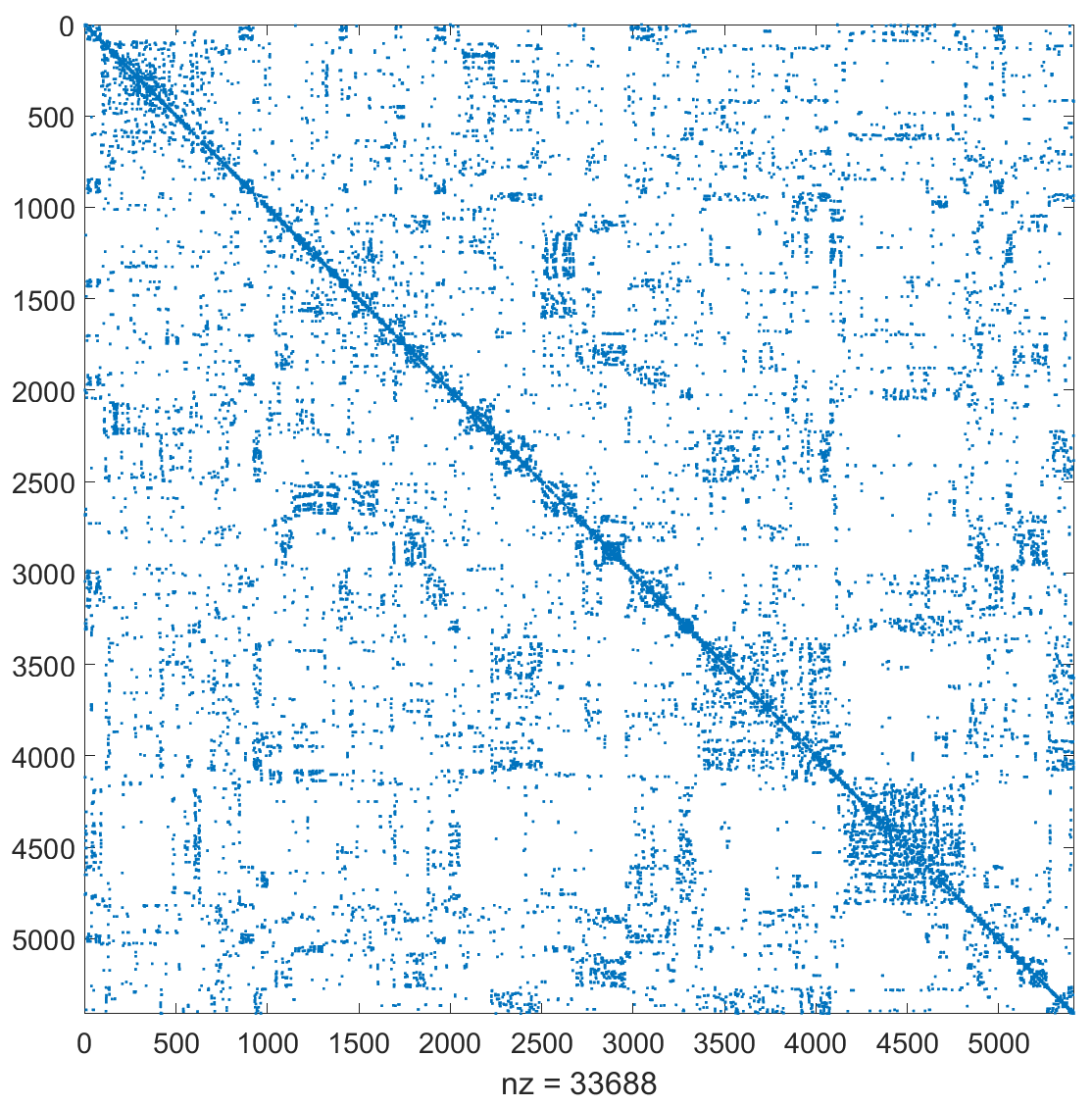}\\
		\multicolumn{3}{c}{Sparsity matrix}\\
		\hline
		a) GOA Province&b) UK &c) NY State
	\end{tabular}
	\caption{Region Adjacency graphs and sparse adjacency matrices. a) Genova Province (GOA Province) partitioned into 137 zones b) United Kingdom (UK) partitioned into 344 local authority districts c) New York State (NY State) partitioned into 5410 census tracts. }\label{figure:sparsity_matrices}
\end{figure}

For instance, the Region Adjacency graphs (\autoref{sec:TRAFFIC-GRAPH}) are undirected graphs that are highly sparse since they represent the geometric partition of a geographical region where most of the units have just a few neighbours. However, there can be some areas where the sub-regions are clustered, which produces square sub-matrices along the diagonal of the sparsity matrix (Fig.~\ref{figure:sparsity_matrices}).

\subsection{Centrality metrics for graphs\label{sec:CENTRALITY-METRICS}}
Different centrality metrics can be defined for the nodes in a directed graph~$\mathcal{G}=(V,E)$ with~$N$ nodes. The main centrality metrics are degree centrality, closeness centrality, harmonic centrality, betweenness centrality, and Page rank centrality. The closeness and harmonic centralities measure the likelihood of a node reaching the rest of the nodes, which can be extended to the concept of being reached if the graph is directed. The betweenness centrality measures the importance of a node to create connections between the rest of the nodes, and it is computed in the same way for undirected and directed graphs. Finally, PageRank centrality is a metric specifically for directed graphs since the influence of a node depends on the influence of the incoming connections.

\begin{figure}[t]
	\centering
	\begin{tabular}{ccc}
		\includegraphics[scale=0.16]{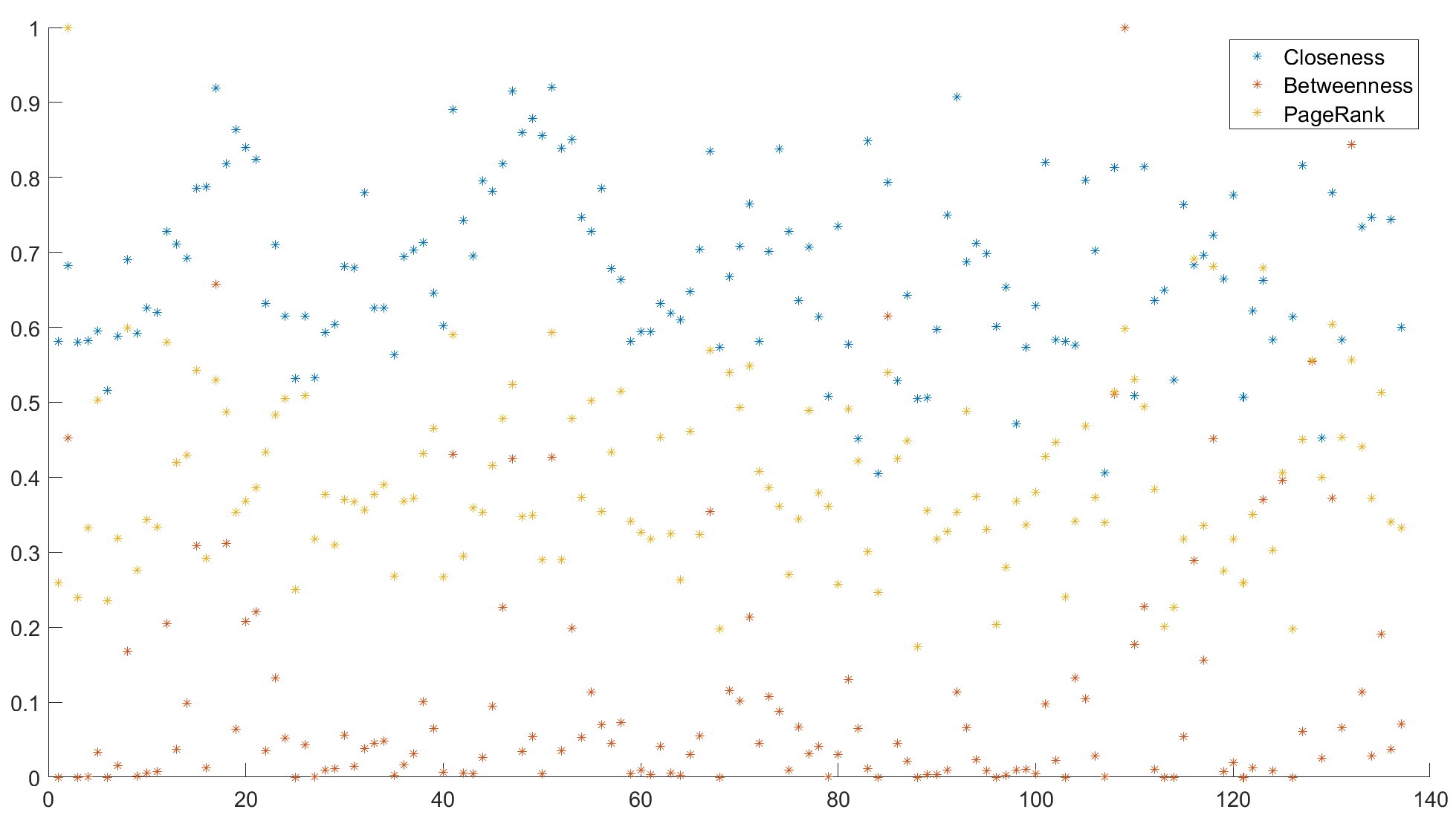}
		&\includegraphics[scale=0.16]{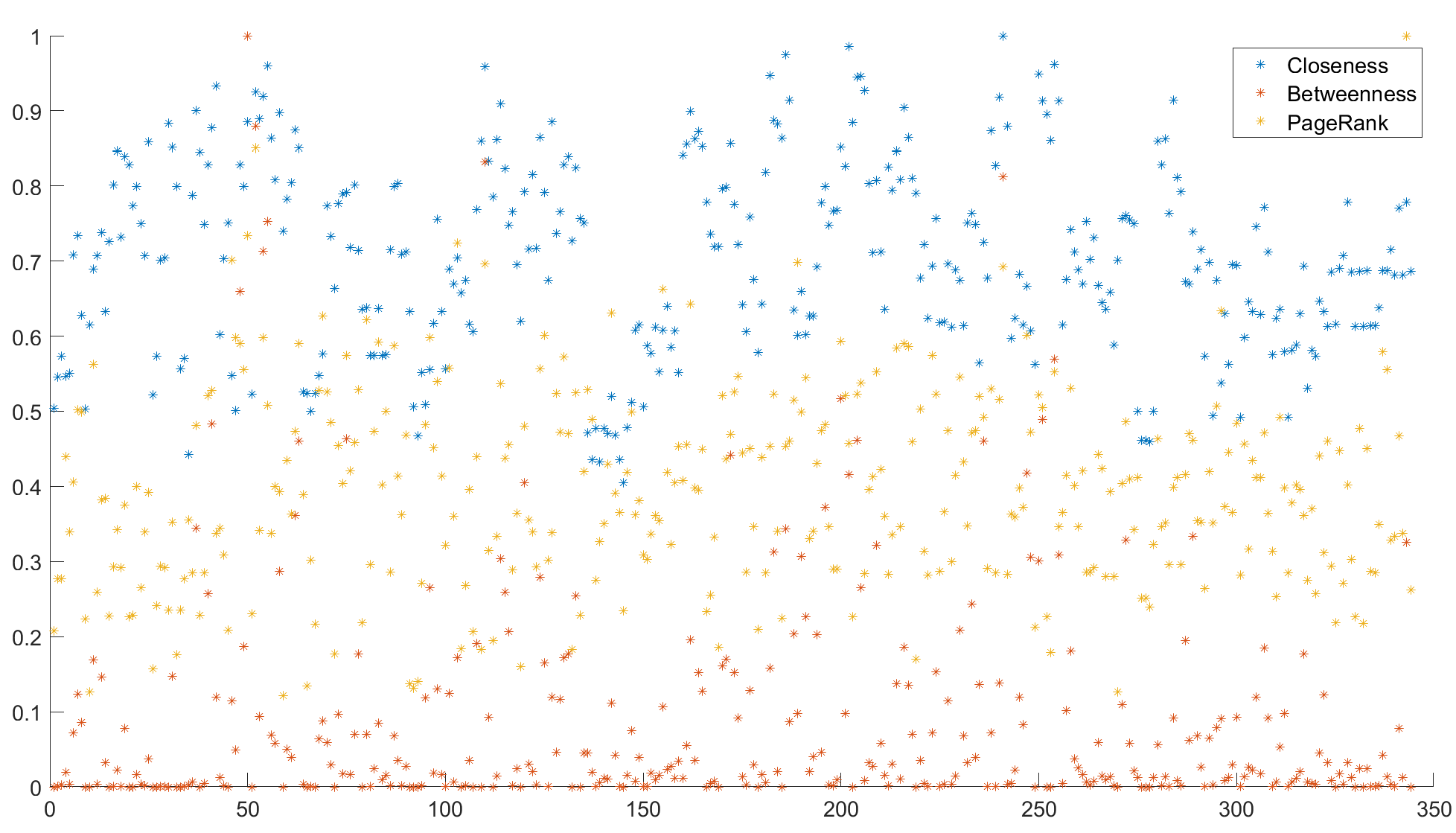}
		&\includegraphics[scale=0.16]{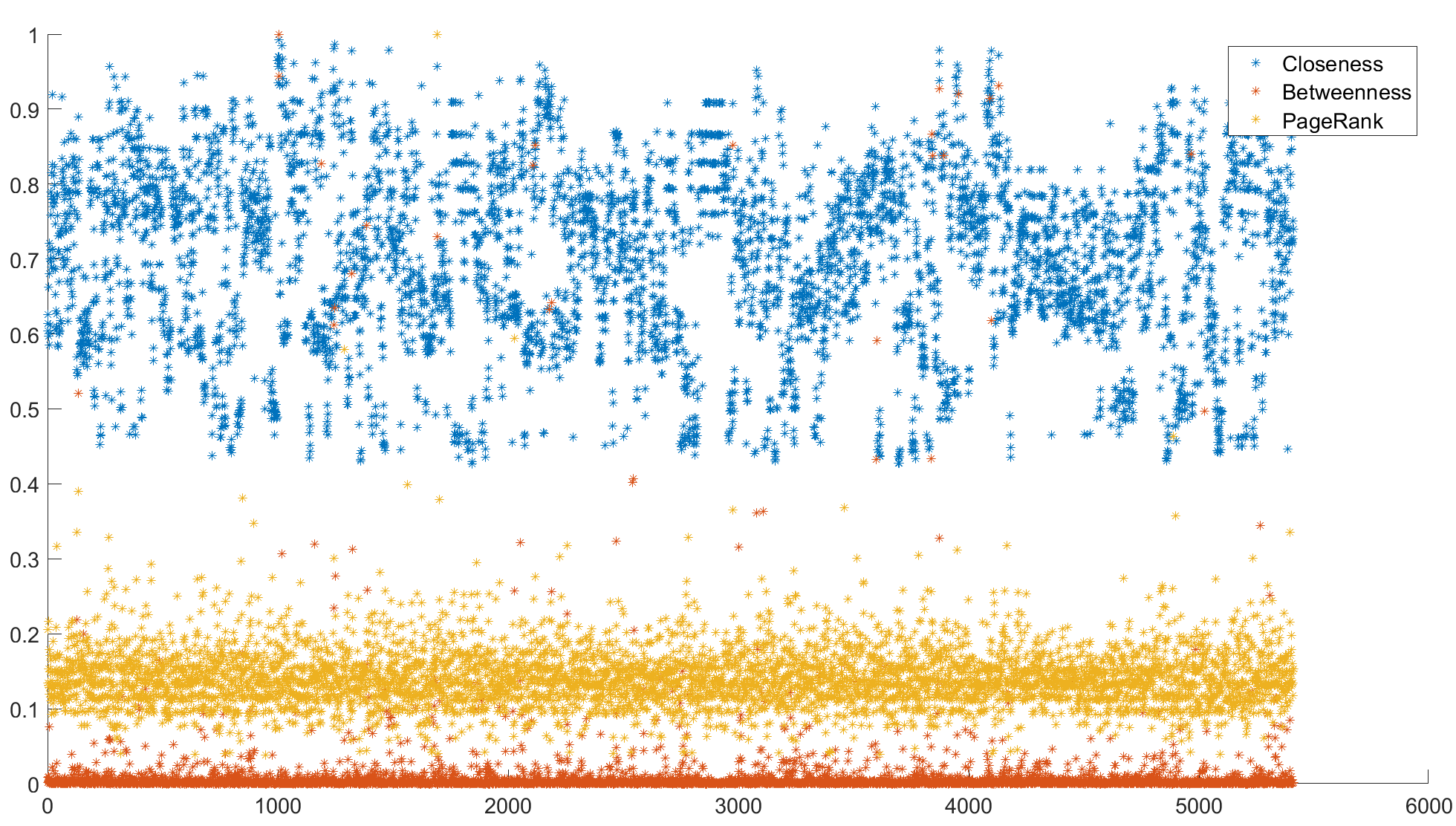}\\
		(a) GOA Province
		&(b) UK 
		&(c) NY
	\end{tabular}
	\caption{Normalised node centrality metrics. The normalised betweenness centrality is the smallest for the three Region Adjacency graphs regardless of the number of nodes. \label{figure:region_adjacency_graphs_normalized_centralities}}
\end{figure}

\paragraph*{Degree centrality:} The \textit{out-degree centrality} and \textit{in-degree centrality} of a node~$v$ in a directed graph are defined by

\begin{equation*}
    D^+(v)=\sum_{j=1}^NA_{vj},\qquad
    D^-(v)=\sum_{j=1}^NA_{jv},
\end{equation*}

respectively. This centrality metric represents how big that node's outgoing or incoming flow is compared to the corresponding flow in the rest of the nodes. 

\paragraph*{Closeness centrality:}
The \textit{out-closeness centrality} and \textit{in-closeness centrality} of a node~$v$ are defined by
\begin{equation*}
    C^+(v)=\dfrac{N-1}{\displaystyle \sum_{i\neq v}d(v,i)},\qquad
    C^-(v)=\dfrac{N-1}{\displaystyle \sum_{i\neq v}d(i,v)}.
\end{equation*}

The two types of closeness centrality defined on a directed graph measure the likelihood of a node reaching the other nodes or being reached by them. For an undirected graph, the closeness centrality measures the likelihood of a node being connected to the rest of the nodes in a graph. If for some nodes~$v,b$ there is no path from~$v$ to~$b$, then we would have~$C(v)=0$. The bigger the distances from~$v\in V$ to the rest of the nodes are, the smaller the value of the closeness centrality~$C(v)$ is, the nodes that are closer to the rest of the graph have higher closeness centrality values. The closeness centrality for a directed graph may have more nodes with zero values, compared to the undirected case, since more cases of not symmetric paths arise, for instance, in a not strongly connected graph.

\paragraph*{Harmonic centrality:}
The \textit{out-harmonic centrality} and \textit{in-harmonic centrality} of a node~$v$ are defined by

\begin{equation*}
    H^+(v)=\dfrac{1}{N-1} \sum_{i\neq v}\frac{1}{d(v,i)},\qquad
    H^-(v)=\dfrac{1}{{N-1}}\sum_{i\neq v}\frac{1}{d(i,v)}.
\end{equation*}

The harmonic centrality in a directed graph reduces zero closeness centrality values when the graph is not strongly connected. It simply excludes the not-reachable nodes to provide positive values for most nodes. A node will have a zero out-harmonic centrality value if and only if its out-degree is zero. A node will have a zero in-harmonic centrality value if and only if its in-degree is zero. From the inequality between the arithmetic mean and the harmonic mean of real values, it follows \mbox{$H^+(v)\geq C^+(v)$} and \mbox{$H^-(v)\geq C^-(v)$}. For any undirected graph~$\mathcal{G}$ and any node~$v\in V$, from the inequality between the arithmetic mean and the harmonic mean of real values it follows \mbox{$H(v)\geq C(v)$} and the harmonic centrality is just a variation of the closeness centrality to handle the infinite values caused by disconnected graphs since the existence of a distance~$d(v,b)=\infty$ would vanish the value of~$C(v)$ discarding the possibility of further analysis for the node~$v$. 

\paragraph*{Betweenness centrality:} Similarly to the undirected case, the \textit{betweenness centrality} of a node~$v$ is defined by

\begin{equation*}
    B(v)=\underset{(a,b)\in \mathcal{P}_v}{\sum} \dfrac{\vert S_v(a,b)\vert}{\vert S(a,b)\vert},
\end{equation*}

where~$S(a,b)$ is the set of shortest paths from~$a$ to~$b$,~$S_v(a,b)=\{w\in S(a,b)\colon v\in w\}$ is the set of shortest paths from~$a$ to~$b$ with~$v$ as an intermediate node,~$\mathcal{P}_v=\{(a,b)\in V\times V\colon a,b\neq v, a\neq b, S(a,b)\neq \varnothing\}\}$ is the set of node couples that have~$v$ as part of a shortest path, and~$v\in w=(v_1,\hdots,v_m)$ means~$v=v_i$ for some~$i=2,\hdots,m-1$. The notation~$\vert X \vert$ represents the cardinality of a set~$X$, in other words, the number of elements of~$X$. The betweenness centrality values for the nodes in a directed graph are generally smaller than the values in the undirected case because~$S(a,b)\neq \varnothing$ does not imply that~$S(b,a)\neq \varnothing$ and therefore there might be fewer elements in the sum.

For directed graphs, the betweenness centrality estimates the importance of a node~$v$ as a connection between the rest of the graph. Namely, it measures the proportion of shortest paths between any other couple of nodes passing through~$v$. The higher the betweenness centrality of a node is, the more significant the proportion of shortest paths between other nodes passing through it. A node with a high betweenness centrality value can be interpreted as a node whose removal would considerably affect the graph since some nodes may remain without an optimal path between them, and some nodes may even become disconnected. If no short paths between any couple of nodes pass through a given (and different) node, then its betweenness centrality is zero. If for every~$a,b\in V$ it holds~$S(a,b)=\varnothing$, then the graph consists of isolated nodes, and the betweenness centrality values can be set to zero for each node.

\paragraph*{PageRank centrality:} This centrality metric is particularly used with directed graphs; it measures the influence of a node~$v\in V$ depending on the influence of every other node~$u\in V$ such that~$(u,v)\in E$ (incoming edges to~$V$). Intuitively, if a node has incoming edges from a relevant node, its influence in the graph would be bigger than if those edges were from a less relevant node. The \textit{PageRank centrality} of a node~$v$ originally defined in~\citep{page_rank_original}, is defined in a recursive way through

\begin{equation*}
    PR(v)=(1-c)+c\left(\dfrac{PR(v_1)}{D^+(v_1)}+\hdots \dfrac{PR(v_m)}{D^+(v_m)}\right),
\end{equation*}

where the~$v_i$ are the nodes in~$\mathcal{G}$ such that there is an edge from~$v_i$ to~$v$, for~$i=1,\hdots,m$. Moreover,~$c\in (0,1)$ is a value called the damping factor that helps to deal with dead-end nodes (without outgoing edges). Usually, it is set to~$c=0.85$. The PageRank values can be seen as the principal eigenvector of a matrix~$\mathbf{\hat{P}}^{\top}$ given by \mbox{$ \mathbf{\hat{P}}=c(\mathbf{P}+\delta \cdot b^{\top})+(1-c)\mathbf{E}$}, where~$\mathbf{P}$ is the transition probability matrix~$\mathbf{P}$ (\autoref{subsection:transitionmatrix}),~$\mathbf{b}\in \mathbb{R}^N$ is a distribution probability called teleportation vector, whose component~$i$ denotes the probability to move arbitrarily to the node~$i$ while moving along the edges of the graph (with direction), and~$\mathbf{E}:=(1,\hdots,1)\mathbf{b}^{\top}$~\citep{page_rank_survey}). The vector~$\delta\in \mathbb{R}^N$ has components~$\delta_i:=\delta (d_i,0)$ (Kronecker delta) for~$i=1,\hdots, N$, where~$d_i$ is the degree of the node~$i$.

\begin{table}[t]
	\centering
	\caption{Summary of centrality metrics for the nodes of a graph}
	\begin{tabular}{c|c|c}
		\textbf{Metric and definition}&\textbf{Undirected graphs}  &\textbf{Directed graphs}\\\hline
		\begin{tabular}{c}\textit{Degree centrality}\\
			$D(v)=\sum_{j=1}^NA_{vj}$
		\end{tabular}&Yes&Out-degree and in-degree variations\\\hline
		\begin{tabular}{c}\textit{Closeness centrality}\\
			~$C(v)=\dfrac{N-1}{\displaystyle \sum_{i\neq v}d(v,i)}.$
		\end{tabular}&Yes &Outgoing and incoming variations\\\hline
		\begin{tabular}{c}\textit{Harmonic centrality}\\
			~$H(v)=\dfrac{\displaystyle \sum_{i\neq v}\frac{1}{d(v,i)}}{N-1}$
		\end{tabular}&Yes &Outgoing and incoming variations\\\hline
		\begin{tabular}{c}\textit{Betweenness centrality}\\
			~$B(v)=\underset{(a,b)\in \mathcal{P}_v}{\sum} \dfrac{\vert S_v(a,b)\vert}{\vert S(a,b)\vert}$
		\end{tabular}&Yes&Yes\\\hline
		\begin{tabular}{c}\textit{PageRank centrality}\\
			~$PR(v)=(1-c)+c\displaystyle \sum_{u\rightarrow v}\dfrac{PR(u)}{d(u)}$
		\end{tabular}&No&Yes\\\hline
	\end{tabular}
\end{table}

For instance, after normalising each centrality metric for the three Region Adjacency graphs from Fig.~\ref{figure:sparsity_matrices}, a trend in common is shown for all of them (Fig.~\ref{figure:region_adjacency_graphs_normalized_centralities}), i.e. the closeness centrality has invariably the largest values. In contrast, the betweenness centrality has the smallest ones because the maximum value of the betweenness centrality is larger than the maximum value of the other two centralities. Moreover, for the NY State graph, the majority of betweenness centrality values are localised in the first decile of its distribution as the number of nodes in the NY State graph is considerably larger than in the other two graphs, which results in larger betweenness centralities because every node is part of the shortest path between more node pairs.

\paragraph*{Flow and circulation of a directed graph}
In a graph~$\mathcal{G}$, we consider a function \mbox{$F:\mathcal{E}(\mathcal{G})\rightarrow\mathbb{R}^{+}$} that assigns to each directed edge \mbox{$(i,j)$} a non-negative value \mbox{$F_{ij}$}, which is said to be a \emph{circulation} if at each node~$i$ we have that \mbox{$\sum_{j,\, j\rightarrow i}F_{ji}=\sum_{j,\, i\rightarrow j}F_{ij}$}. A circulation is said to be \emph{invertible} if \mbox{$F_{ij}=F_{ji}$}. For a strongly connected directed graph~$\mathcal{G}$, the eigenvector~$\phi$ of the transition probability matrix~$\mathbf{P}$ with eigenvalue~$1$ is associated with a circulation

\begin{equation}
	\label{equation_circulation_perron}
F^{\phi}_{ij}:=\phi_iP_{ij}.
\end{equation}

\begin{figure}[t]
	\centering
	\begin{tabular}{ccc}
		\multicolumn{3}{c}{Undirected Region Adjacency graphs}\\
		\includegraphics[scale=0.19]{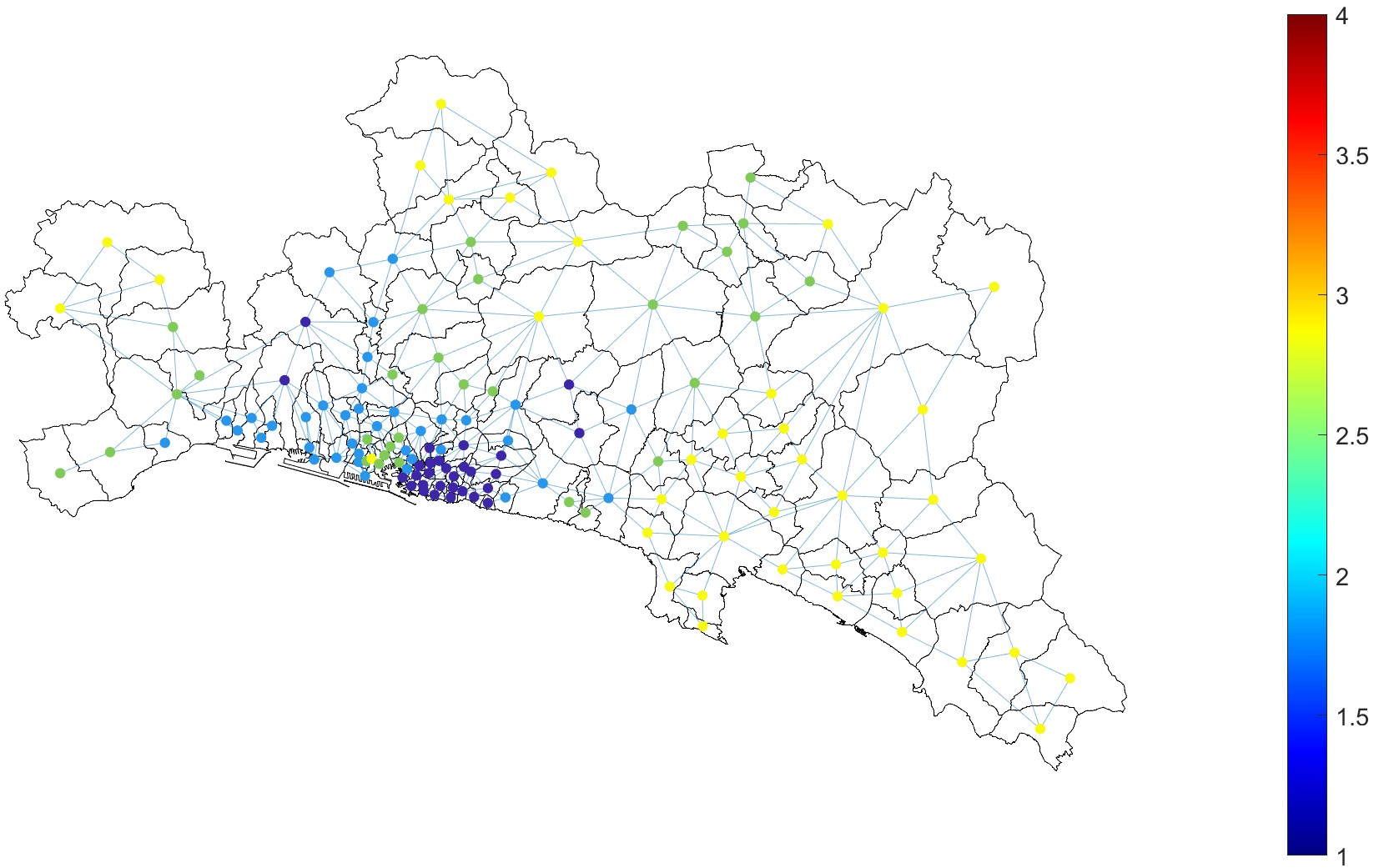}
		&\includegraphics[scale=0.19]{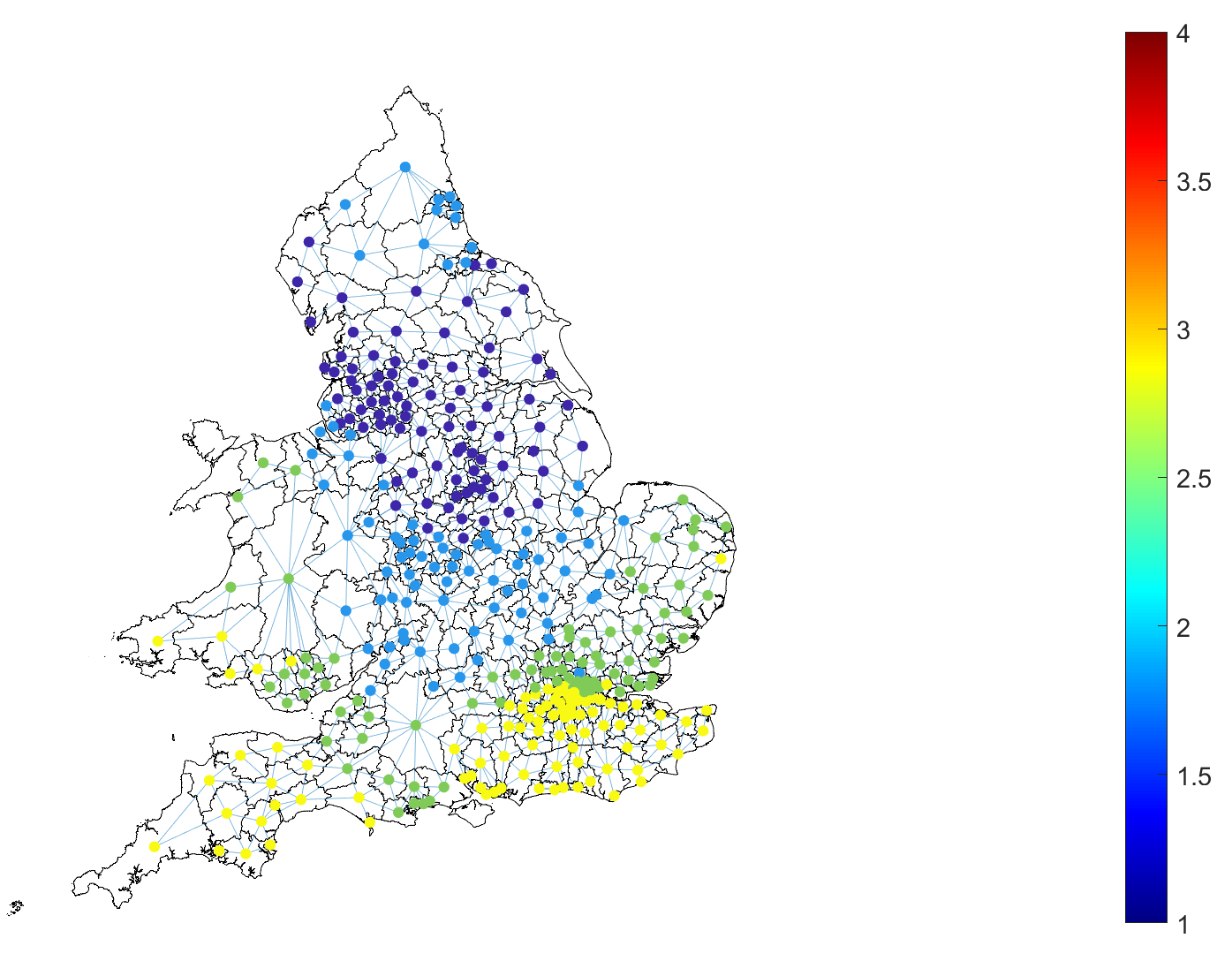}&\includegraphics[scale=0.19]{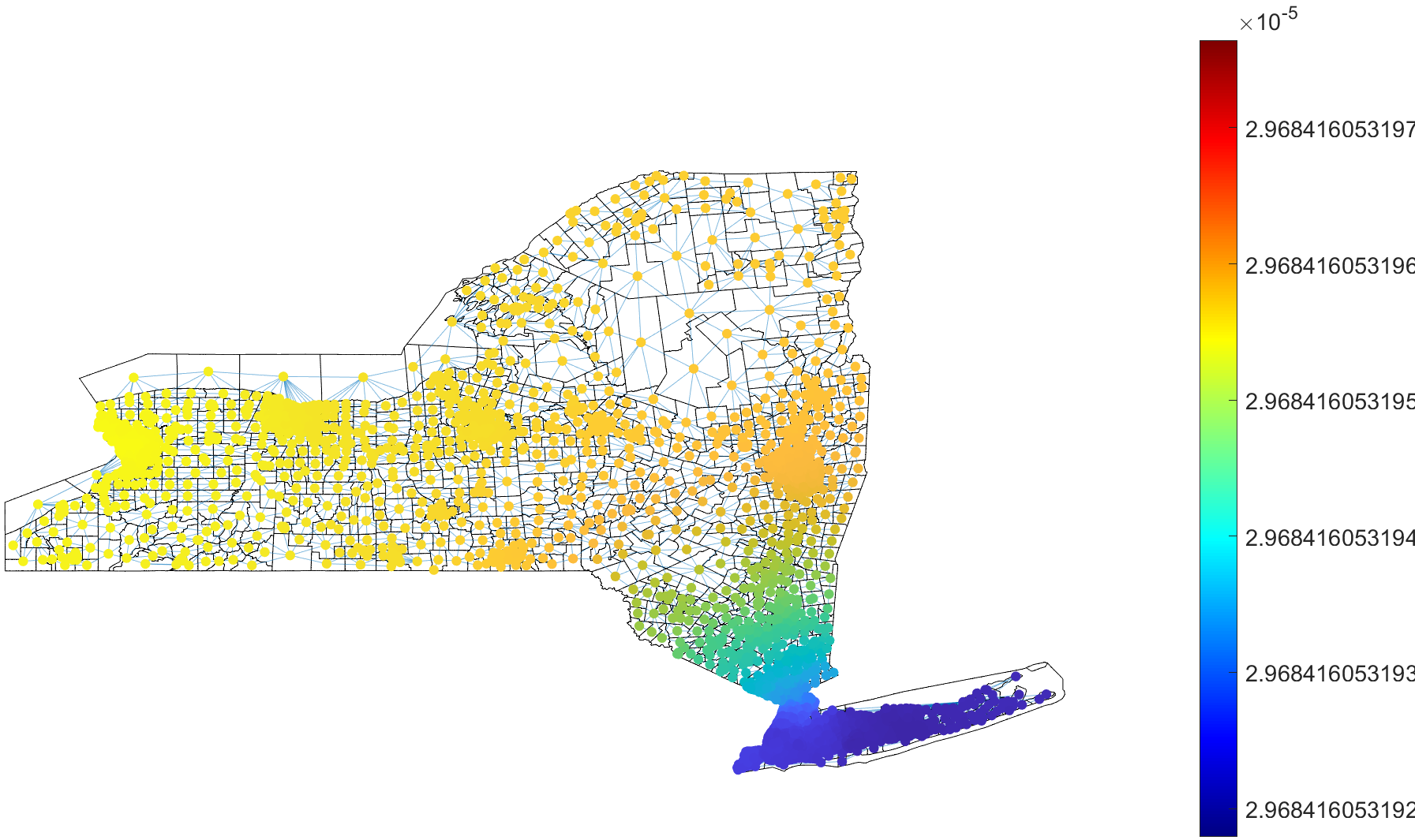}\\
		(a) GOA Province &(b) UK &(c) NY State
	\end{tabular}
	\caption{Average node circulation values for Region Adjacency graphs. The values for a) GOA Province and b) UK correspond to the quartiles partition to enhance the visualisation. \label{figure:region_adjacency_graphs_perron_circulation_normalised_NY}}
\end{figure}

The \textit{average node circulation} of each node is defined by~$\Tilde{F^{\phi}_i}=\sum_{j}F^{\phi}_{ij}/D^+(i)$. For instance, as the Region-Adjacency graphs are connected (after possibly removing the "islands", i.e. nodes with node degree equal to~$1$), it is possible to define a circulation associated to their Perron vector by using Eq.~\ref{equation_circulation_perron}. The trend of the average node circulation (Fig.~\ref{figure:region_adjacency_graphs_perron_circulation_normalised_NY}) has an inverse behaviour with respect to the closeness centrality values in Fig.~\ref{figure:region_adjacency_graphs_closeness}. Namely, the nodes with higher average circulation have the lowest closeness centrality values, i.e., the nodes with less accessibility to the rest of the network. This inverse behaviour is more evident for the GOA Province from the central subregions to the eastern ones. The inverse behaviour is clearer for the UK from the central subregions to the southern ones. Since the number of nodes of the NY State Region Adjacency graph is considerably larger than the other two graphs, the average node circulation values variation is also larger, so its visualisation does not require a quartile partition. 

\subsection{Traffic graphs and metrics\label{sec:TRAFFIC-GRAPH}}

\paragraph*{Region Adjacency and Origin-Destination graphs}
We now introduce two types of traffic analysis and simulation graphs: the Region Adjacency graph and the Origin-Destination graph. Given a finite non-empty compact partition~$\mathcal{P}$ of a region~$K\subset \mathbb{R}^2$, i.e. a collection~$\{R_1,\hdots,R_N\}$ of non-empty compact subsets of~$\mathbb{R}^2$ such that~$\bigcup_i R_i=K$, we can identify each element~$R_i$ of~$\mathcal{P}$ with its centroid~$c_i=(x_i,y_i)\in \mathbb{R}^2$. The \textit{Region Adjacency graph} associated to~$\mathcal{P}$ consists of the undirected unweighted graph with~$N$ nodes embedded in~$V=\{c_1,\hdots,c_N\}$, and edges~$E=\{(c_i,c_j)\colon \partial R_i\cap \partial R_j\neq \varnothing\}$, where~$\partial X$ represents the boundary of the set~$X$.

\begin{figure}[t]
	\centering
	\begin{tabular}{cc}
		\multicolumn{2}{c}{Directed Origin-Destination graph}\\
		\includegraphics[height=150pt]{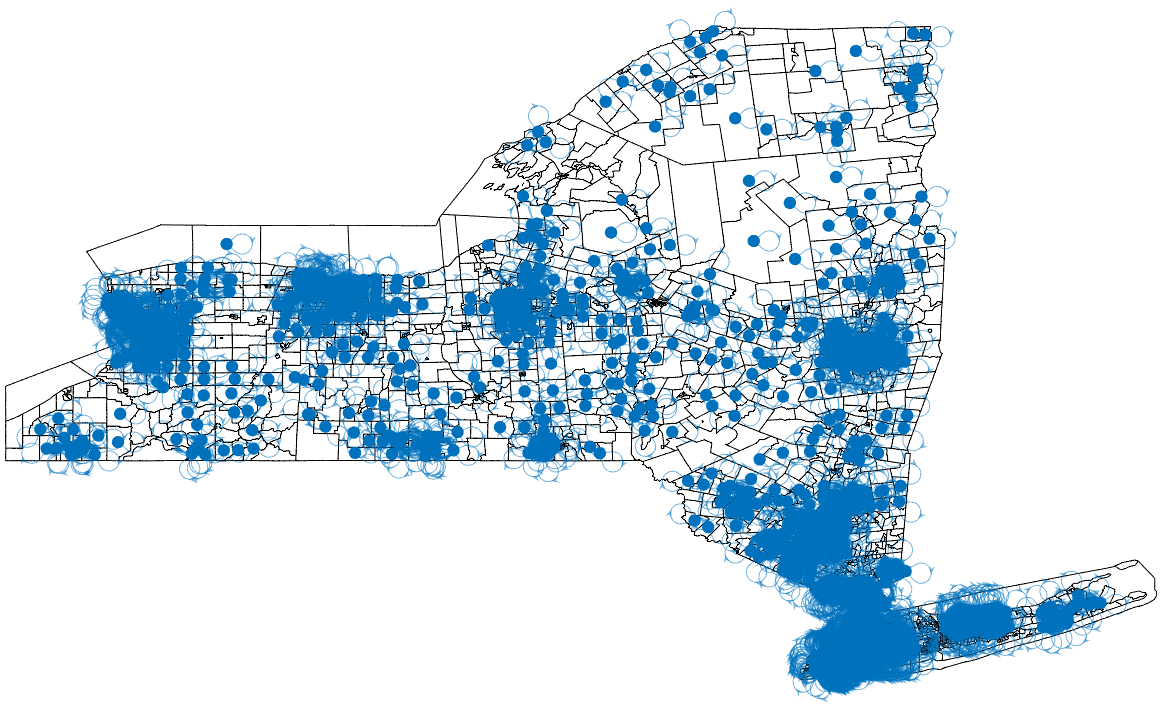}
		&\includegraphics[height=150pt]{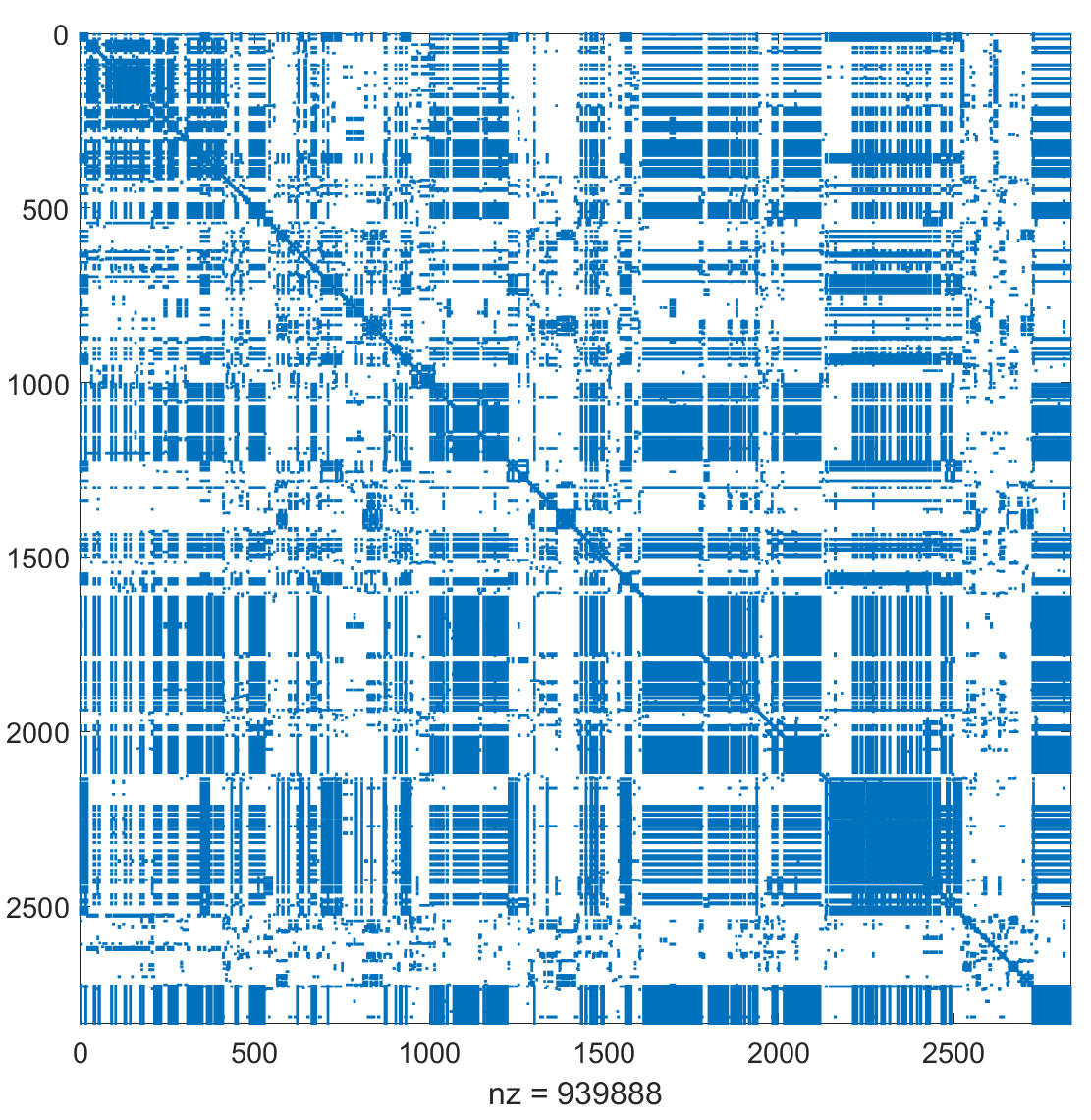}\\
		(a) NY State Origin-Destination graph &(b) Sparsity matrix
	\end{tabular}
	\caption{Origin-Destination digraph associated to the predicted OD matrix for NY State using the Deep Gravity mobility model.\label{figure:NY_OD_digraph}}
\end{figure}

Given an Origin-Destination matrix~$M\in \mathbb{R}^{N\times N}$, the \textit{Origin-Destination graph} associated to~$M$ consists of the directed weighted graph with weighted adjacency matrix~$M$. The~$N$ nodes can be embedded in some~$\mathbb{R}^2$ representation associated to~$M$. For instance, we can construct an Origin-Destination graph for the New York State using the predicted flows generated by the Deep Gravity mobility model (\autoref{section:deepgravity}). The entry~$M_{ij}$ of the weighted adjacency matrix~$M$ is the predicted flow from the location indexed by~$i$ to the location indexed by~$j$. The Origin-Destination graph~$\mathcal{G}$ associated to~$M$ (Fig.~\ref{figure:NY_OD_digraph}) consists of~$2836$ nodes and~$939888$ edges, and it is not connected since the Deep Gravity model predicts flows only for the locations in the same regular partition, i.e. the same square cell.

The adjacency matrices for the Region Adjacency graphs are highly sparse (Fig.~\ref{figure:sparsity_matrices}). In contrast, the Origin-Destination graph shows more clusters with a more significant number of connections since the Deep Gravity mobility model predicts flows for some regions that are not adjacent geographically (Fig.~\ref{figure:NY_OD_digraph}a), thus its adjacency matrix (Fig.~\ref{figure:NY_OD_digraph}b) has many more square sub-matrices that are related to the number of different strongly connected components since the Deep Gravity model predicts flows only for the irregular units within the same squared cell.

\begin{figure}[t]
	\centering
	\begin{tabular}{ccc}
		\multicolumn{3}{c}{Undirected Region Adjacency graphs}\\
		\hline
		\multicolumn{3}{c}{\ }\\
		\includegraphics[scale=0.20]{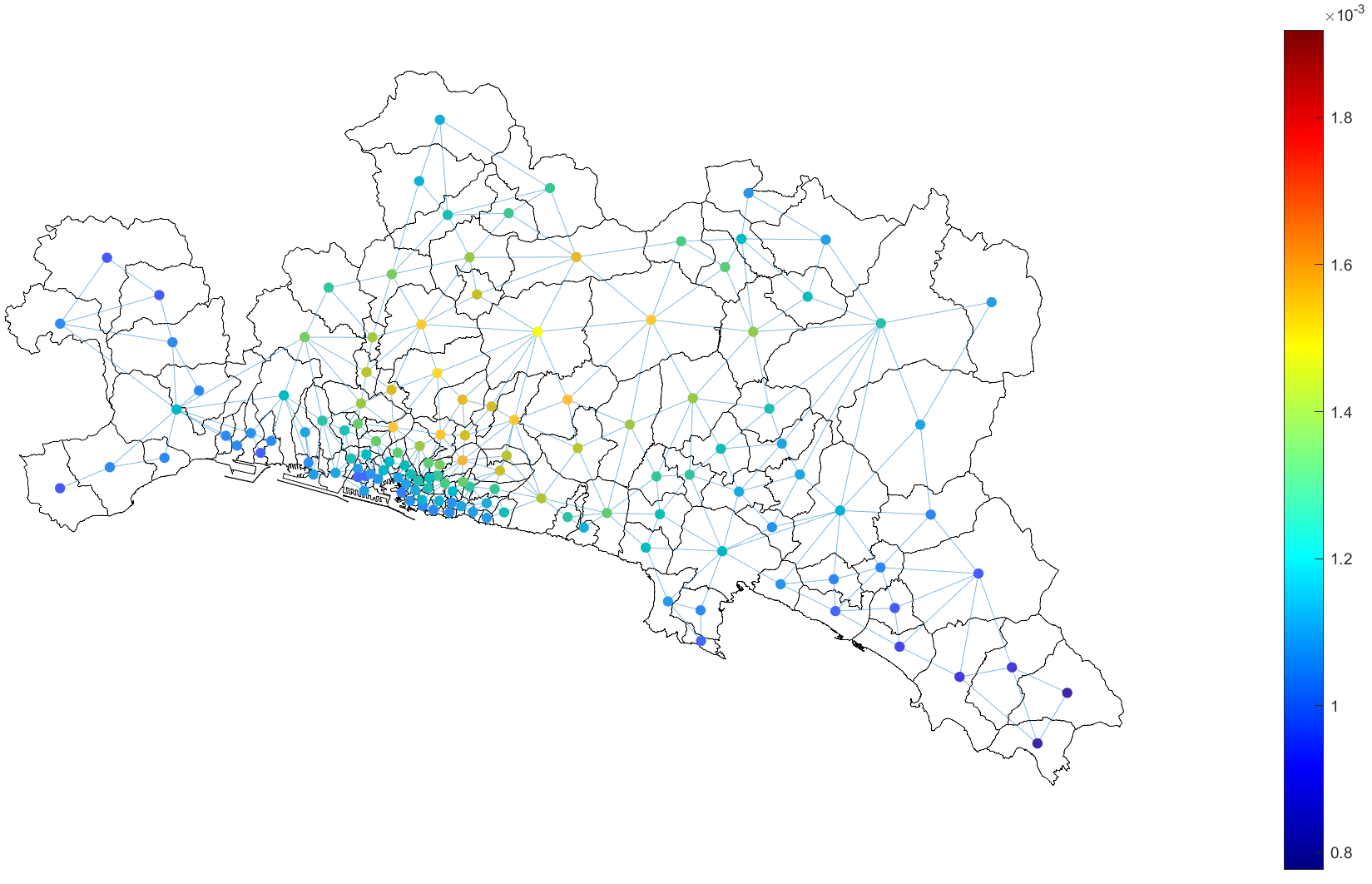}
		&\includegraphics[scale=0.20]{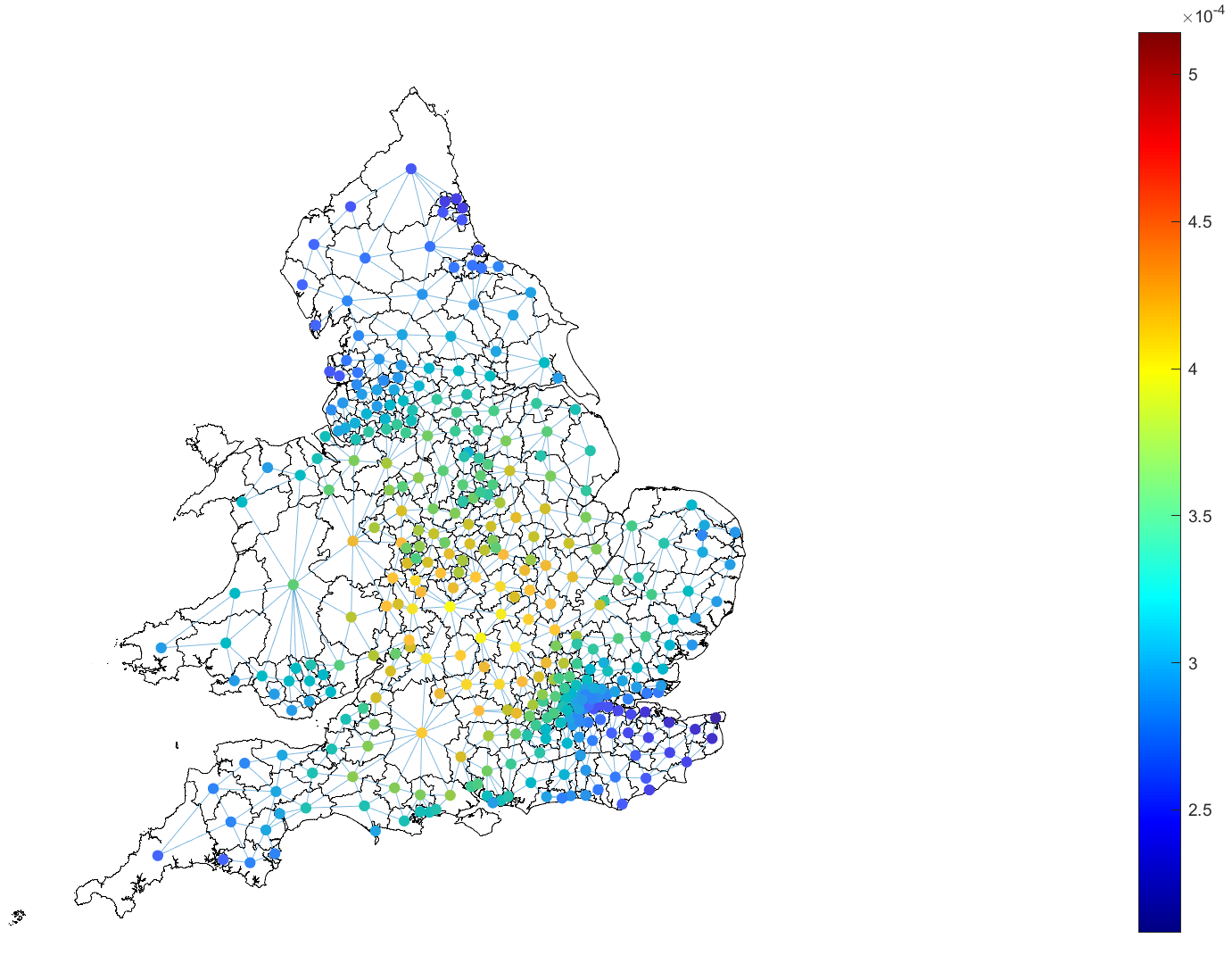}
		&\includegraphics[scale=0.20]{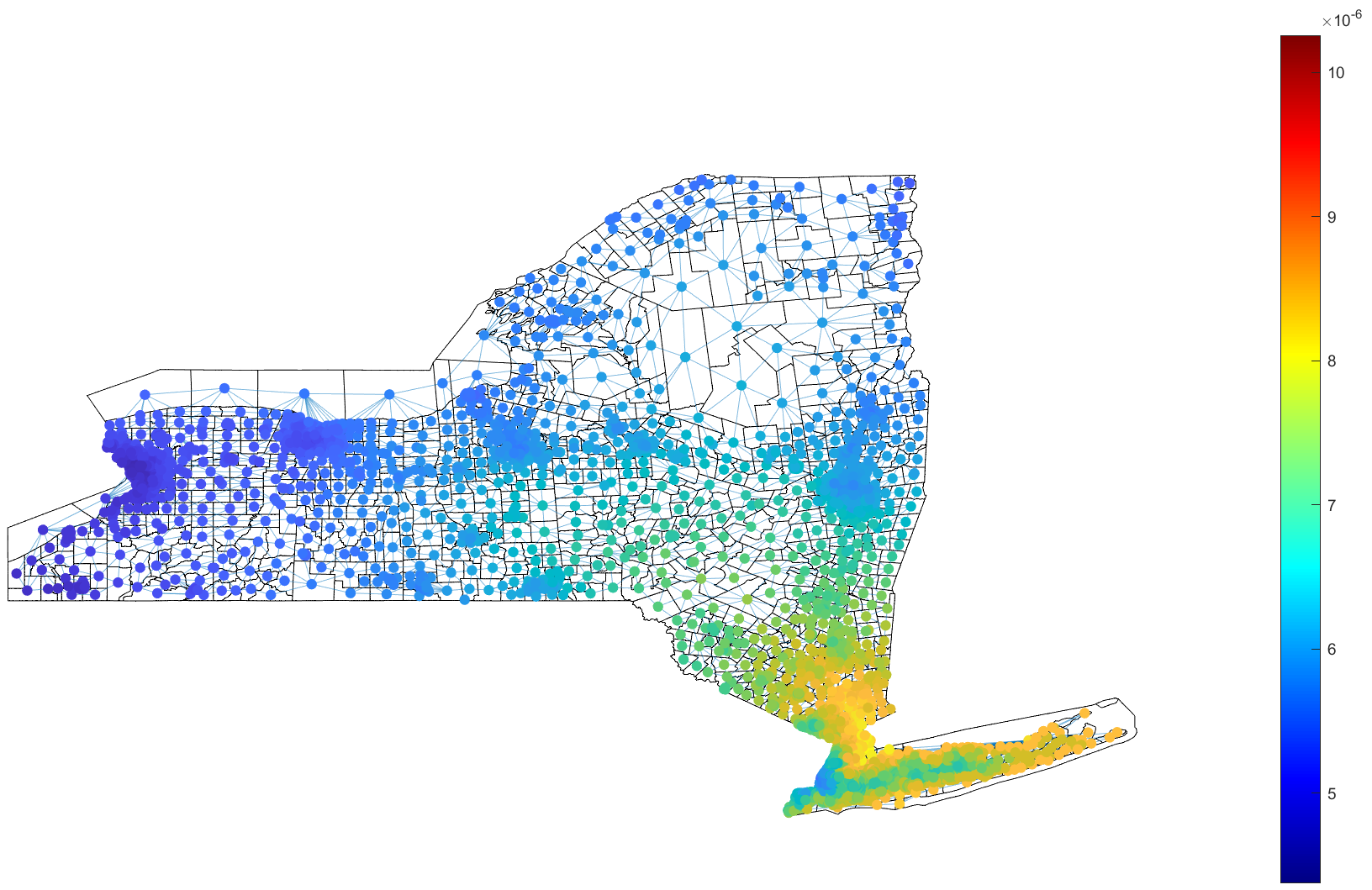}\\
		\multicolumn{3}{c}{Closeness centrality}\\
		\hline
		\multicolumn{3}{c}{\ }\\
		\includegraphics[scale=0.20]{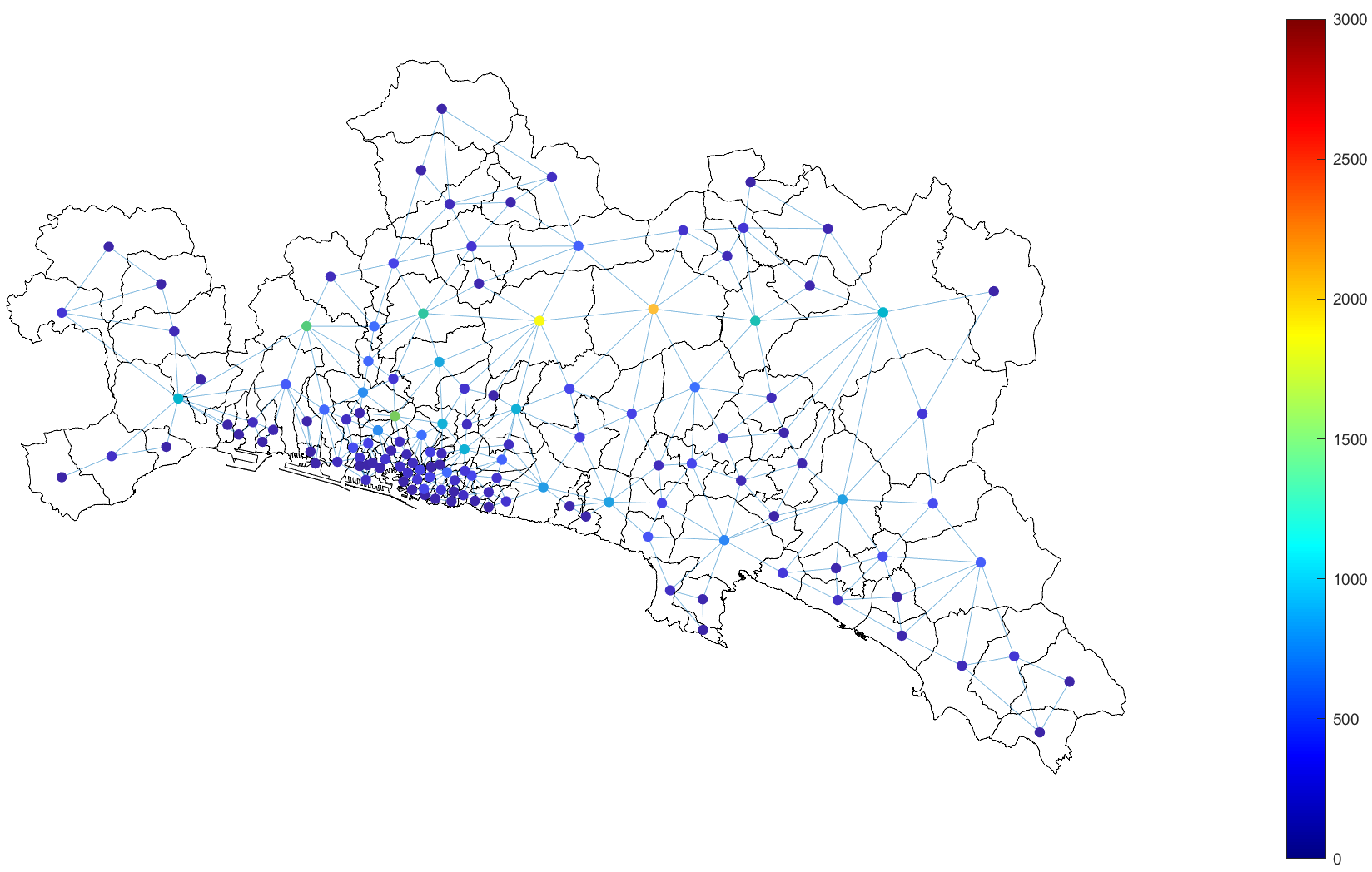}
		&\includegraphics[scale=0.20]{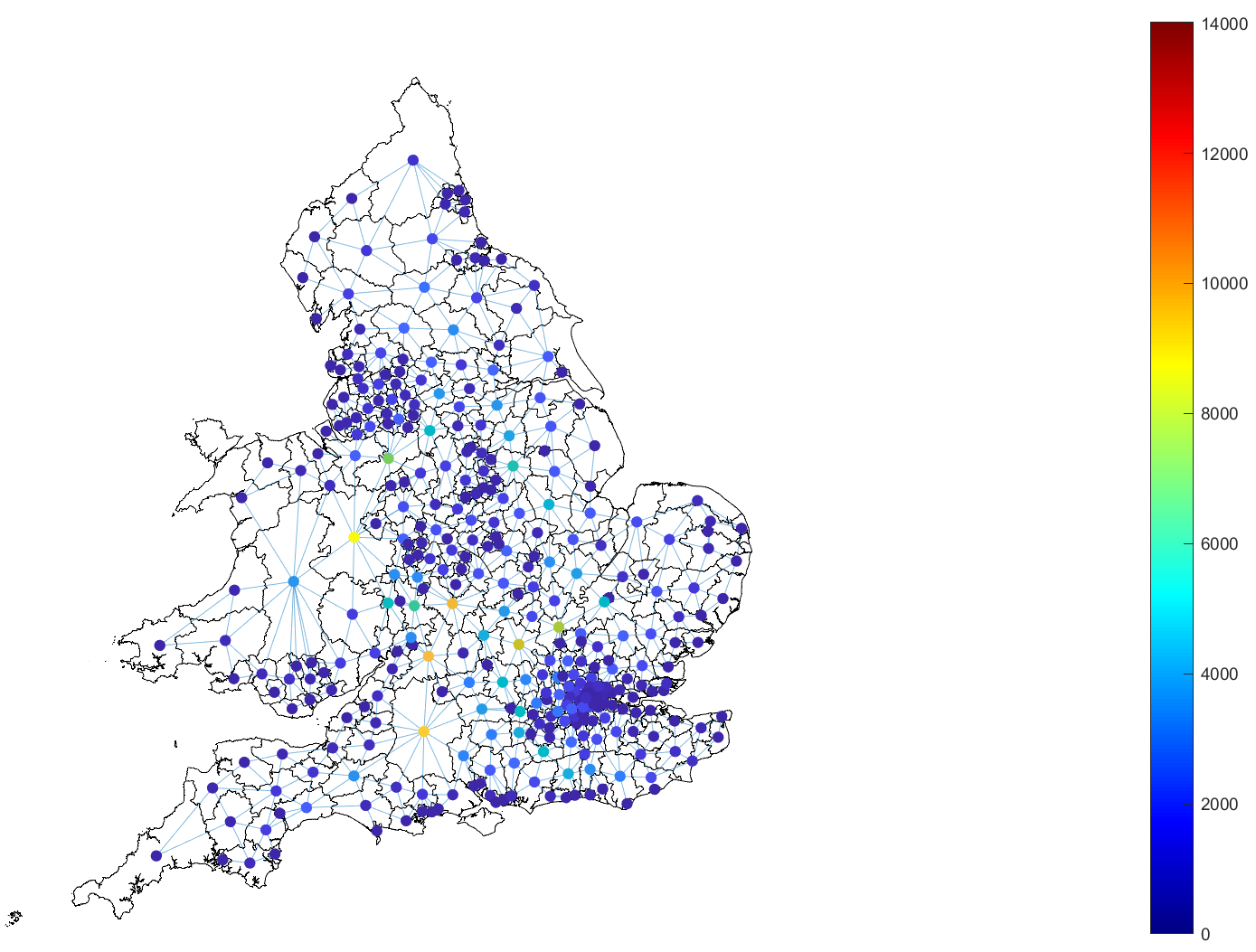}
		&\includegraphics[scale=0.20]{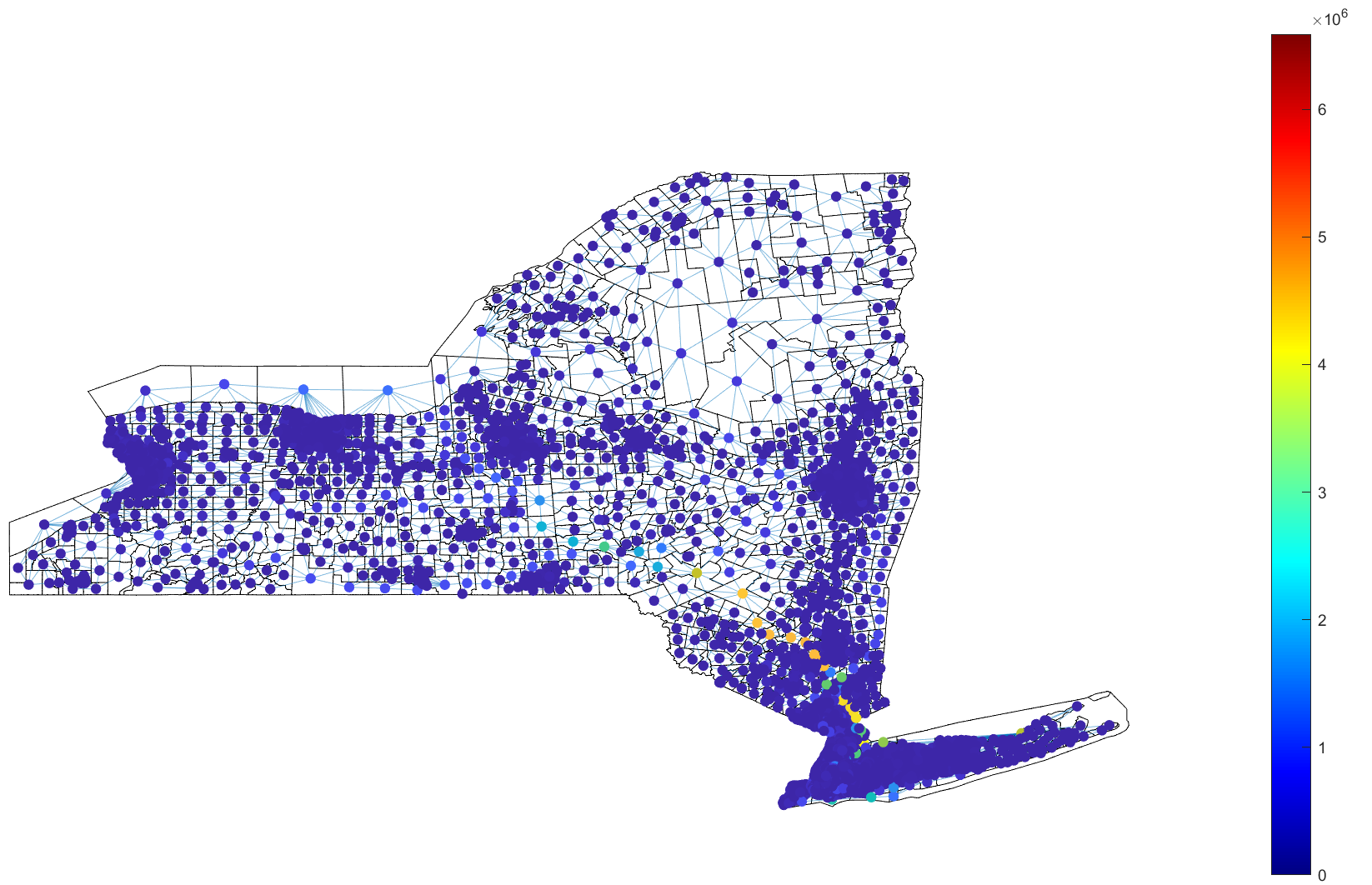}\\
		\multicolumn{3}{c}{Betweenness centrality}\\
		\hline
		(a) GOA Province&(b) UK&(c) NY State
	\end{tabular}
	\caption{Closeness and Betweenness centralities for Region Adjacency graphs. The closeness values show significant changes depending on their geographical position, while the betweenness values are nearly constant.
		\label{figure:region_adjacency_graphs_closeness}}
\end{figure}

\paragraph{Metrics on the Region Adjacency graph} The closeness centrality values of the nodes in a Region Adjacency graph can have a clustered behaviour showing the subregions that have higher levels of accessibility to the rest of the subregions in a given planar partition. For instance,  there is a trend of lower levels of accessibility from the peripherical subregions. In contrast, the central subregions have higher closeness centrality values (Fig.~\ref{figure:region_adjacency_graphs_closeness}), except for NY State, which has a significant higher values cluster in the Southeast area, which could be due to its geometrical shape and the number of subregions in that area. The betweenness centrality values in the same Region Adjacency graph are, in general, low for most of the network (Fig.~\ref{figure:region_adjacency_graphs_closeness}), so there are no clusters of subregions with higher connectivity importance that serve to link other subregions through the shortest path. Nevertheless, in the NY State Region Adjacency graph, there are a few nodes with the highest betweenness centrality values around the "bottleneck" that connects the Southeast area to the rest of the network, which means that their removal may have an impact on the shortest paths between the subregions in the Southeast area and the ones in the rest of the NY State. Indeed, this difference between the maximum betweenness centrality values and the values in the rest of the nodes are the ones that cause the low trend for the normalised betweenness centrality (Fig.~\ref{figure:region_adjacency_graphs_normalized_centralities}).

\begin{figure}[t]
	\centering
	\begin{tabular}{ccc}
		\multicolumn{3}{c}{Directed Origin-Destination graph}\\
		\includegraphics[scale=0.20]{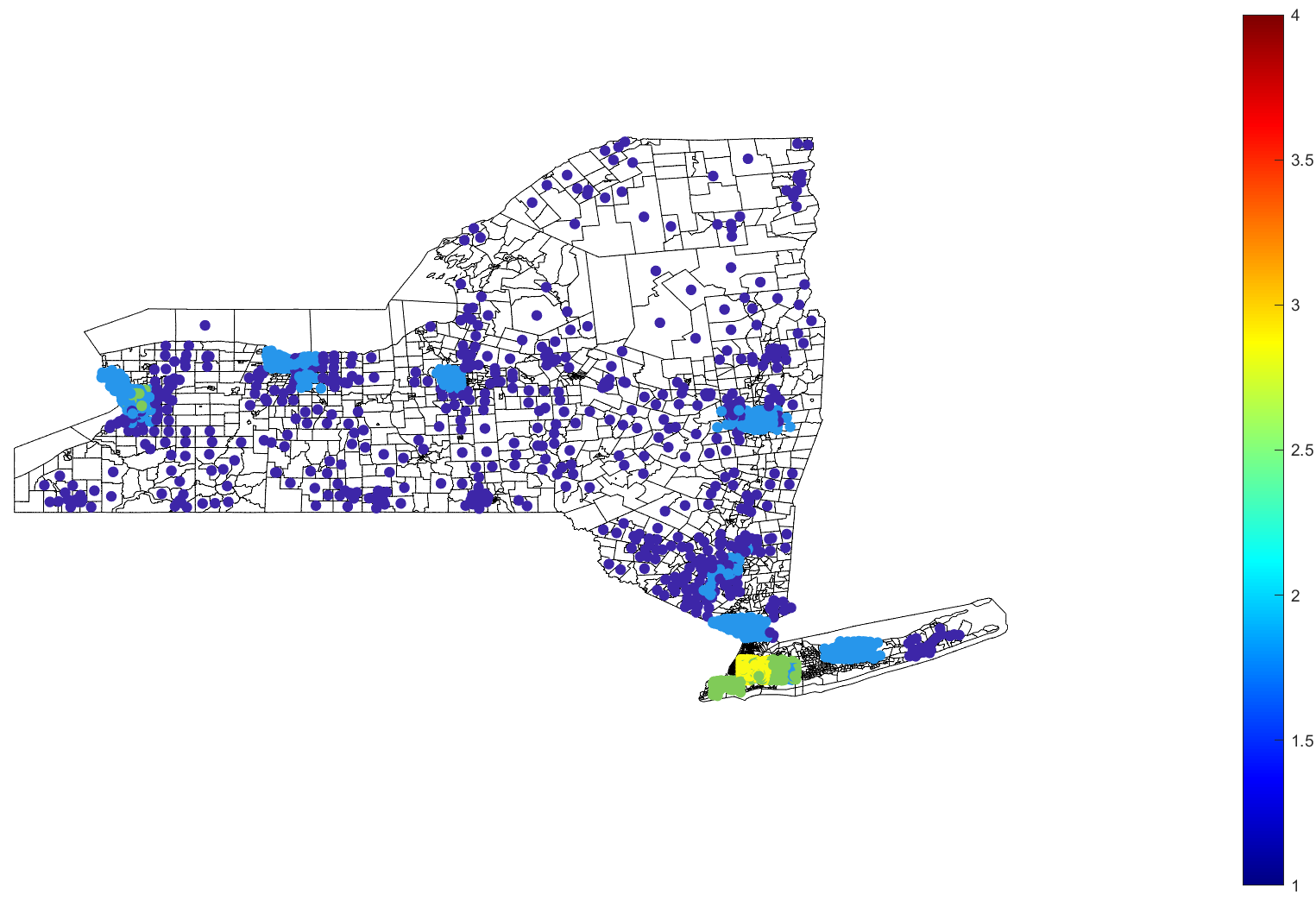}
		&\includegraphics[scale=0.20]{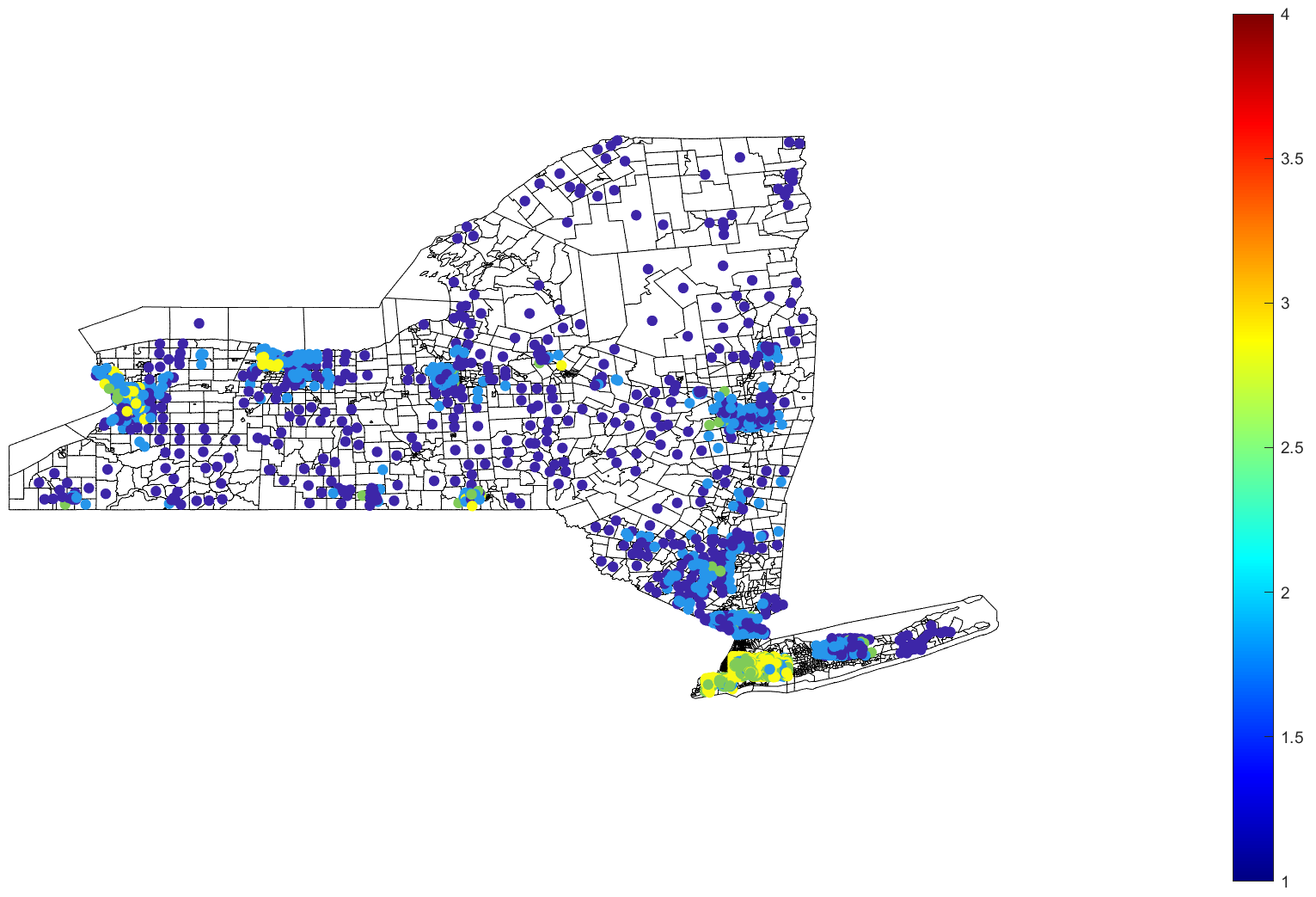}
		&\includegraphics[scale=0.20]{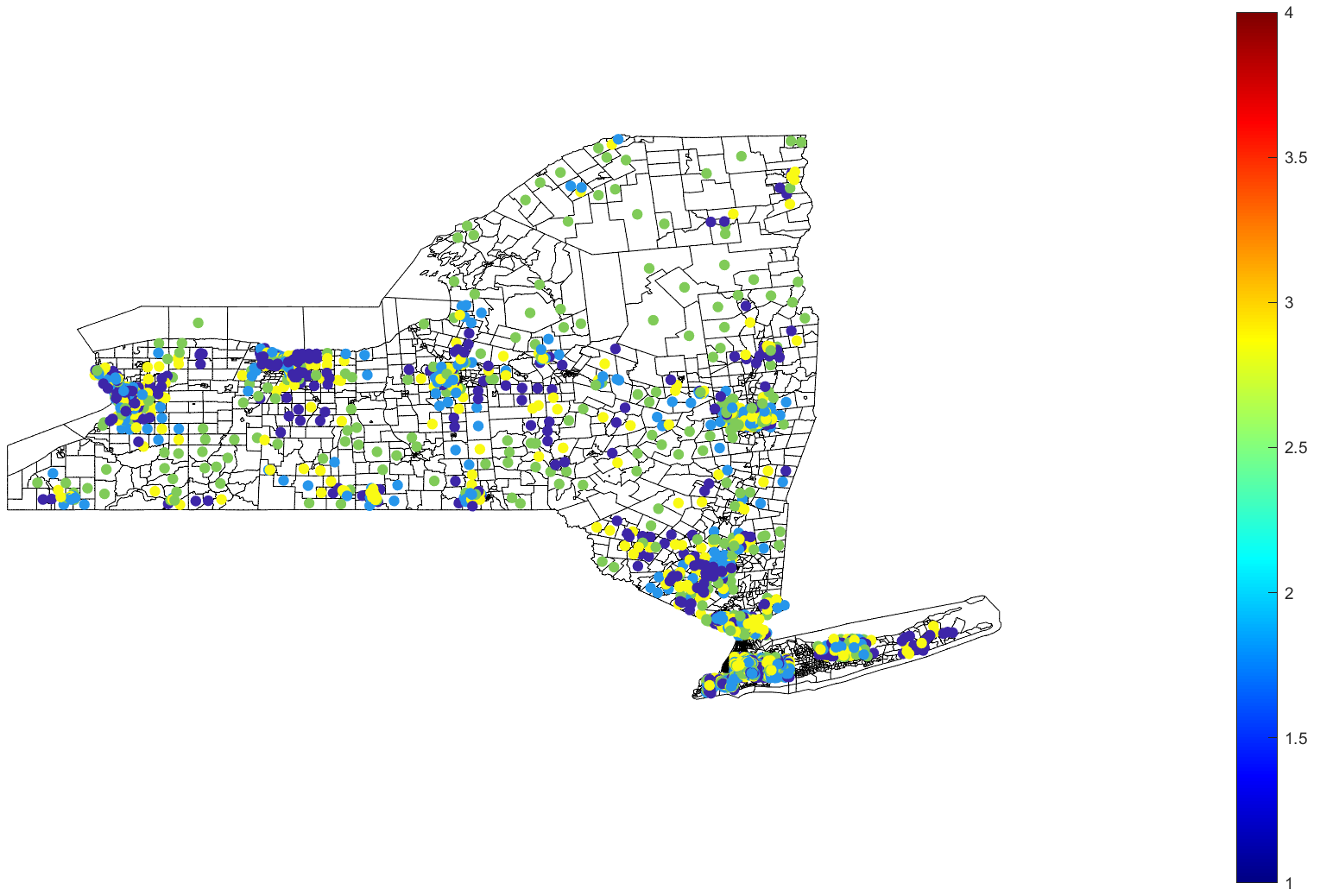}\\
		(a) Outcloseness centrality
		&(b) Betweenness centrality
		&(c) Page rank centrality
	\end{tabular}
	\caption{Quartiles visualisation of Centrality metrics for the NY State Origin-Destination graph. The edges are not shown for enhancing the visualisation of node centrality values. \label{figure:ny_od_centrality_values}}
\end{figure}

\paragraph{Metrics on Origin-Destination graphs} As with the Region Adjacency graphs, the centrality metrics values of the nodes in an Origin-Destination graph can also reveal some information regarding the connectivity of the subregions in a given planar partition, and their flows represented in an Origin-Destination matrix. However, it is possible that the available flows are not sufficient to generate a strongly connected graph. Indeed, the Deep Gravity mobility model predicts flows only for the irregular subregions within the same squared cell, creating as many strongly connected components as cells exist. The higher values of the outcloseness and betweenness centralities of the Origin-Destination graph for New York State (Fig.~\ref{figure:ny_od_centrality_values}) are localised approximately in the same subregions as the nodes with higher closeness centrality values in the Region Adjacency graph of New York State (Fig.~\ref{figure:region_adjacency_graphs_closeness}). This trend occurs because the Origin-Destination graph of New York State has significantly more nodes in this area, creating a strongly connected component with a larger number of nodes and consequently with more accessible nodes and connections, which increases outcloseness and betweenness centralities, respectively. The Page rank centrality exhibits a more randomised trend because several strongly connected components have various nodes. The original values were classified in quartiles to visualise their distribution better.

\section{Graph matrices for network and traffic analysis\label{sec:GRAPH-MATRICES}}
The node degree centrality motivates the definition of the degree matrix, which associates a value to each node based on the number of its neighbours. The transition probability matrix represents the connection's relevance for each node's neighbours, as its entries are values that measure the proportion of information flow across the edges of the graph. Furthermore, the \textit{graph Laplacian} allows us to study structure properties, such as connectedness through eigenvalue problems, and there are different types of Laplacians on a graph~$\mathcal{G}$, depending on the model represented by~$\mathcal{G}$, and also whether the graph is undirected or directed. For connected graphs and strongly connected graphs, the transition probability matrix has an associated \textit{Perron vector} that is used to define several Laplacian matrices and a circulation function that provides another sense of information flow between any couple of nodes of the graph.

\subsection{Degree matrix\label{subsection:adjacency}}

If~$\mathcal{G}$ is a graph with~$N$ nodes (weighted or unweighted, directed or undirected), then the \textit{degree} of~$v\in V$ is defined as the sum of the elements of the~$v$-th row of the adjacency matrix. Namely, the degree~$d_v$ of a node~$v\in V$ is defined by \mbox{$d_v=\sum_{j=1}^NA_{vj}$}. Indeed, the degree of a node is the sum of the weights of the edges joining that node for undirected graphs and the sum of the weights of the edges outgoing from that node for digraphs. The \textit{degree matrix}~$\mathbf{D}$ is the diagonal matrix~$\mathbf{D}\in \mathbb{R}^{N\times N}$ given by \mbox{$D_{ij}=d_i$} if \mbox{$i=j$} and~$0$ otherwise.

\subsection{Transition probability matrix\label{subsection:transitionmatrix}}
Let~$\mathcal{G}$ be a weighted graph and let~$i\in V$ be a node with~$d_i>0$ outgoing edges (arriving to the nodes~$v_1,\hdots,v_{d_i}$). The proportion of the flow at node~$i$ that will continue through node~$k$ is given by~$p_{ik}=A_{ik}/d_i$, for~$k=v_1,\hdots,v_{d_i}$. On the other hand, if~$d_i=0$ then the proportion of outgoing flows from~$i$ to~$k$ is simply zero for every~$k\in V$. This discussion motivates the definition of the \textit{transition probability matrix}~$\mathbf{P}$, which defines a \textit{Markov chain} associated with random walks on the graph~$\mathcal{G}$~\citep{diplacian_yanhua}, where the entry~$P_{ij}$ denotes the probability of moving from node~$i$ to node~$j$, for~$i,j=1,\hdots,N$. However, it is required that there are no nodes without any edge for undirected graphs or nodes without outgoing edges (called \textit{dead-end} nodes, possibly with incoming edges, for directed graphs). The reason is that if~$i\in V$ is a node with such conditions. The proportion of the flow going from~$i$ to~$j$ will be zero for any node~$j\in V$; therefore, the whole~$i$-th row would be zero. Still, a transition matrix of a Markov chain must satisfy having the sum of each column equal to~$1$ (\textit{row stochastic}).

For the sake of theoretical completeness in Markov chains~\citep{diplacian_yanhua}, we assume that a graph is strongly connected to define its transition probability matrix. The \textit{transition probability matrix}~$\mathbf{P}$ of the graph~$\mathcal{G}$ is the~$N\times N$ matrix that satisfies the relation \mbox{$\mathbf{P}=\mathbf{D}^{-1}\mathbf{A}$}, where~$\mathbf{A}$ and~$\mathbf{D}$ are the adjacency and degree matrices, respectively. The entries~$p_{ij}$ represent the probability of the flow moving from node~$i$ to node~$j$ (or the proportion of flows that moves from~$i$ to~$j$), for~$i,j=1,\hdots,N$. The matrix~$\mathbf{P}$ is sometimes called the \textit{normalised adjacency matrix} of~$\mathcal{G}$~\citep{primerlaplacian_veerman}. From the definition of the degree matrix~$\mathbf{D}$ it follows that

\begin{equation*}
    \sum_{j=1}^NP_{ij}=\sum_{j=1}^N(\mathbf{D}^{-1}\mathbf{A})_{ij}=\sum_{j=1}^N\sum_{k=1}^ND^{-1}_{ik}A_{kj}=\sum_{j=1}^N\dfrac{1}{d_i}A_{ij}=\dfrac{1}{d_i}\sum_{j=1}^NA_{ij}=1,
\end{equation*}

for every~$i=1,\hdots,N$, in other words, the sum of all the entries of a row equals~$1$, for every row in~$\mathbf{P}$. Indeed, the matrix~$\mathbf{P}$ is row stochastic, and there is a Markov chain associated with random walks on the graph~$\mathcal{G}$ defined by~$\mathbf{P}$.

\paragraph*{Perron vector}
 For a strongly connected directed graph, the transition probability matrix~$\mathbf{P}$ has a unique left eigenvector~$\phi$ with positive components, namely, \mbox{$\phi^{\top}\mathbf{P}=\rho \phi^{\top}$}. The vector~$\phi$ is called the \textit{Perron vector} of~$\mathbf{P}$, and in fact, it can be easily proven that~$\rho=1$. We notice that~$\phi$ is the Perron vector of the transition probability matrix \mbox{$\mathbf{P}:=\mathbf{D}^{-1}\mathbf{A}$} if and only if \mbox{$\widetilde{\phi}:=\mathbf{D}^{-1}\phi$} is the generalised eigenvector of the couple \mbox{$(\mathbf{A},\mathbf{D})$} associated with the eigenvalue~$1$, i.e., \mbox{$\mathbf{A}\widetilde{\phi}=\mathbf{D}\widetilde{\phi}$}. Indeed, we solve the generalised eigenproblem and then compute  \mbox{$\phi:=\mathbf{D}^{-1}\widetilde{\phi}$}. Since the eigenvector associated with the eigenvalue~$1$ is unique (as the graph is strongly connected), its entries are all positive or negative; if negative, we change their sign to guarantee that the entries of~$\phi$ are all positive. Since \mbox{$\mathbf{P}\mathbf{1}=\mathbf{1}$}, we get that \mbox{$\rho=1$} is an eigenvalue of~$\mathbf{P}$ and all the eigenvalues of~$\mathbf{P}$ are lower than~$1$. In particular, we normalise the entries of~$\phi$ such that \mbox{$\sum_{i=1}^{n}\phi(i)=1$}.

\subsection{Graph Laplacians\label{sec:GRAPH-LAPLACIAN}}
An essential property of the graph Laplacians of an undirected graph is the symmetry, which follows from the symmetry of the adjacency graph and guarantees that all its eigenvalues are real numbers. The Combinatorial Laplacian and the Normalised Laplacian are two possible definitions when the graph is undirected, being the latter a transformation of the former that results in an upper bound for its real eigenvalues. For a directed graph, it is possible to define the Combinatorial Directed Laplacian, Symmetrized Laplacian, and the Combinatorial Symmetrized Laplacian, which have real eigenvalues even if the adjacency matrix is not symmetric. The Diplacian is another viable definition that involves obtaining complex eigenvalues, which, however, coincides with the Normalised Laplacian when the graph is undirected.
\subsubsection{Laplacians of undirected graphs\label{sec:UNDIRECTED-LAPLACIAN}}
For an undirected graph~$\mathcal{G}$ with~$N$ nodes the \textit{Combinatorial Laplacian}~\citep{primerlaplacian_veerman}, is the~$N\times N$ matrix defined by

\begin{equation}\label{equation_laplacian_combinatorial}
\mathbf{L}=\mathbf{D}-\mathbf{A}.
\end{equation}

Since the adjacency matrix of an undirected graph is symmetric, the Combinatorial Laplacian is also symmetric. Moreover,~$\mathbf{D}$ is positive definite, and it is possible to define a \textit{scalar product} in~$\mathbb{R}^N$ by letting \mbox{$\langle \mathbf{x},\mathbf{y}\rangle_{\mathbf{D}}:= \mathbf{x}^{\top}\mathbf{Dy}$}, for \mbox{$\mathbf{x},\mathbf{y}\in \mathbb{R}^N$}. The Combinatorial Laplacian is also called the \textit{Kirchhoff matrix} of the graph~\citep{kernellaplacian_caughman}.~\citep{gacan_graph_zhang} defines the \textit{normalised graph Laplacian}~$\mathbf{\hat{L}}$ by \mbox{$\mathbf{\hat{L}}=\mathbf{I}-\mathbf{D}^{-1/2}\mathbf{A}\mathbf{D}^{-1/2}$}, where~$\mathbf{I}$ is the~$N\times N$ identity matrix, being~$N$ the number of nodes in the graph. The normalised graph Laplacian of an undirected graph is also symmetric, and it is also positive semidefinite since~$\mathbf{D}$ is an invertible diagonal matrix with positive entries. Furthermore,~$\mathbf{\hat{L}}=\mathbf{D}^{-1/2}\mathbf{LD}^{-1/2}$.

\subsubsection{Laplacians of directed graphs\label{sec:DIRECTED-LAPLACIAN}}
Since the Combinatorial Laplacian of a directed graph is not necessarily symmetric,~\citep{graph_signal_real_time_prediction} defines the \textit{Combinatorial Directed Laplacian} of a digraph~$\mathcal{G}=(V,E,\mathbf{A})$ by

\begin{equation*}
    \mathbf{L_G}=\dfrac{1}{2}(\mathbf{D}_{out}+\mathbf{D}_{in}-\mathbf{A}-\mathbf{A}^{\top}),\quad
    \mathbf{D}_{out}=\sum_{j=1}^NA_{ij},\quad
    \mathbf{D}_{in}=\sum_{j=1}^NA_{ji},
\end{equation*}

where~$\mathbf{D}_{out},\mathbf{D}_{in}$ are the out-degree and in-degree and~$\mathbf{A}$ is the adjacency  matrix. The Combinatorial Directed Laplacian is symmetric regardless of~$\mathcal{G}$ being directed or not. Moreover, if~$\mathcal{G}$ is undirected then \mbox{$\mathbf{L}=\mathbf{L_G}$}. The Combinatorial Directed Laplacian is also positive semi-definite since it can be seen as the Combinatorial Laplacian of an undirected with adjacency matrix \mbox{$\tilde{\mathbf{A}}=(\mathbf{A}+\mathbf{A}^{\top})/2$}. Assuming that the input graph is strongly connected, the unique \textit{stationary probability distribution}~$\mathbf{\phi}$ of the transition probability matrix~$\mathbf{P}$ is defined as the unique vector~$\phi \in \mathbb{R}^N$ with strictly positive components such that \mbox{$\phi^{\top}\mathbf{P}=\phi^{\top}$}. In~\citep{laplacians_chung}, the \textit{Symmetrized Laplacian} of a strongly connected graph~$\mathcal{G}$ is

\begin{equation*}
    \mathbf{\mathcal{L}}=\mathbf{I}-\dfrac{\mathbf{\Phi}^{1/2}\mathbf{P}\mathbf{\Phi}^{-1/2}+\mathbf{\Phi}^{-1/2}\mathbf{P}^{\top}\mathbf{\Phi}^{1/2}}{2},
\end{equation*}

where~$\mathbf{\Phi}=\textrm{diag}(\phi_i)$. The Symmetrised Laplacian is indeed symmetric.~\citep{laplacians_chung} also defines the \textit{Combinatorial Symmetrized Laplacian} by

\begin{equation*}
    \mathbf{\mathcal{L}_G}=\mathbf{\Phi}-\dfrac{\mathbf{\Phi P}+\mathbf{P}^{\top}\mathbf{\Phi}}{2},
\end{equation*}

which is symmetric and positive semi-definite since it can be written as the Combinatorial Laplacian of an undirected graph~$\tilde{G}=(V,\tilde{E},\tilde{A})$ with adjacency matrix \mbox{$\tilde{\mathbf{A}}=(\mathbf{\Phi P}+\mathbf{P^{\top}\Phi})/2$}. The Combinatorial Symmetrised Laplacian coincides with the Combinatorial Laplacian defined in~\eqref{equation_laplacian_combinatorial} when the graph is undirected. However,~$\mathbf{\mathcal{L}}$ does not capture the unique characteristic of random walks on digraphs, since different directed graphs can have the same~$\mathbf{\mathcal{L}}$. To overcome this problem,~\citep{diplacian_yanhua} defines the \textit{Diplacian}~$\mathbf{\Gamma}$ though

\begin{equation*}
\mathbf{\Gamma}=\mathbf{\Phi}^{1/2}(\mathbf{I}-\mathbf{P})\mathbf{\Phi}^{-1/2},
\end{equation*}

for which the strongly connected assumption for the graph~$\mathcal{G}$ still holds since the stationary probabilities are required.

\subsubsection{Properties and discussion/comparison\label{sec:GRAPH-PROPERTIES}}
The Laplacian matrix associated with a graph is an essential operator for network models because it constitutes the foundation for Deep Learning techniques on graph structures. The symmetry of the Laplacian is a desired feature even if the graph is directed. Consequently, it is necessary to define more Laplacian matrices in addition to the classic one in Eq.~\eqref{equation_laplacian_combinatorial}. Some matrices, such as the Combinatorial Directed Laplacian, the Symmetrised Laplacian, and the Combinatorial Symmetrised Laplacian, are always symmetric independently of whether the graph is directed or undirected. Moreover, these Laplacians are also positive semi-definite operators because they could be seen as the Combinatorial Laplacian of an undirected graph~$\tilde{G}=(V,\tilde{E})$ with an appropriate choice for the adjacency matrix~$\tilde{\mathbf{A}}$, so the positive semi-definite property of~$\mathbf{L}$ is inherited to the new matrices.

\begin{table}[t]
    \centering
    \caption{Summary of the properties of different Laplacians}
    {\small{
    \begin{tabular}{|c|c|c|}
    \hline
\textbf{Graph Laplacian}&\textbf{Real Eigenvalues}&\textbf{Positive Semidefinite} \\\hline
\multicolumn{3}{c}{\textbf{Undirected graph}}\\\hline
         \begin{tabular}{c}\textit{Combinatorial Laplacian}\\
       ~$\mathbf{L}=\mathbf{D}-\mathbf{A}$\end{tabular}&Yes&Yes\\\hline
         \begin{tabular}{c}
              \textit{Normalised Laplacian}  \\
             ~$\hat{\mathbf{L}}=\mathbf{I}-\mathbf{D}^{-1/2}\mathbf{AD}^{-1/2}$
\end{tabular}&Yes&Yes\\\hline
\multicolumn{3}{c}{\textbf{Directed graph}}\\\hline
         \begin{tabular}{c}
              \textit{Combinatorial Directed Laplacian}  \\
             ~$\mathbf{L}_G=\frac{1}{2}(\mathbf{D}_{out}+\mathbf{D}_{in}-\mathbf{A}-\mathbf{A}^{\top})$
         \end{tabular}&Yes&Yes\\\hline
         \begin{tabular}{c}\textit{Symmetrized Laplacian}\\
       ~$\mathbf{\mathcal{L}}=\mathbf{I}-\dfrac{\mathbf{\Phi}^{1/2}\mathbf{P\Phi}^{-1/2}+\mathbf{\Phi}^{-1/2}\mathbf{P}^{\top}\mathbf{\Phi}^{1/2}}{2}$\end{tabular}&Yes&Yes
         \\\hline
         \begin{tabular}{c}
              \textit{Combinatorial Symmetrized Laplacian}\\
            ~$\mathbf{\mathcal{L}}_G=\mathbf{\Phi}-\dfrac{\mathbf{\Phi} \mathbf{P}+\mathbf{P}^{\top}\mathbf{\Phi}}{2}$
         \end{tabular}&Yes&Yes\\\hline
         \begin{tabular}{c}
         \textit{Diplacian}\\
       ~$\mathbf{\Gamma}=\mathbf{\Phi}^{1/2}(\mathbf{I}-\mathbf{P})\mathbf{\Phi}^{-1/2}$
         \end{tabular}
         &
              No  
              &-\\\hline
             
    \end{tabular}}}
\end{table}

\begin{figure}[t]
\centering
\begin{tabular}{ccc}
\multicolumn{3}{c}{Undirected Region Adjacency graphs}\\
\includegraphics[scale=0.16]{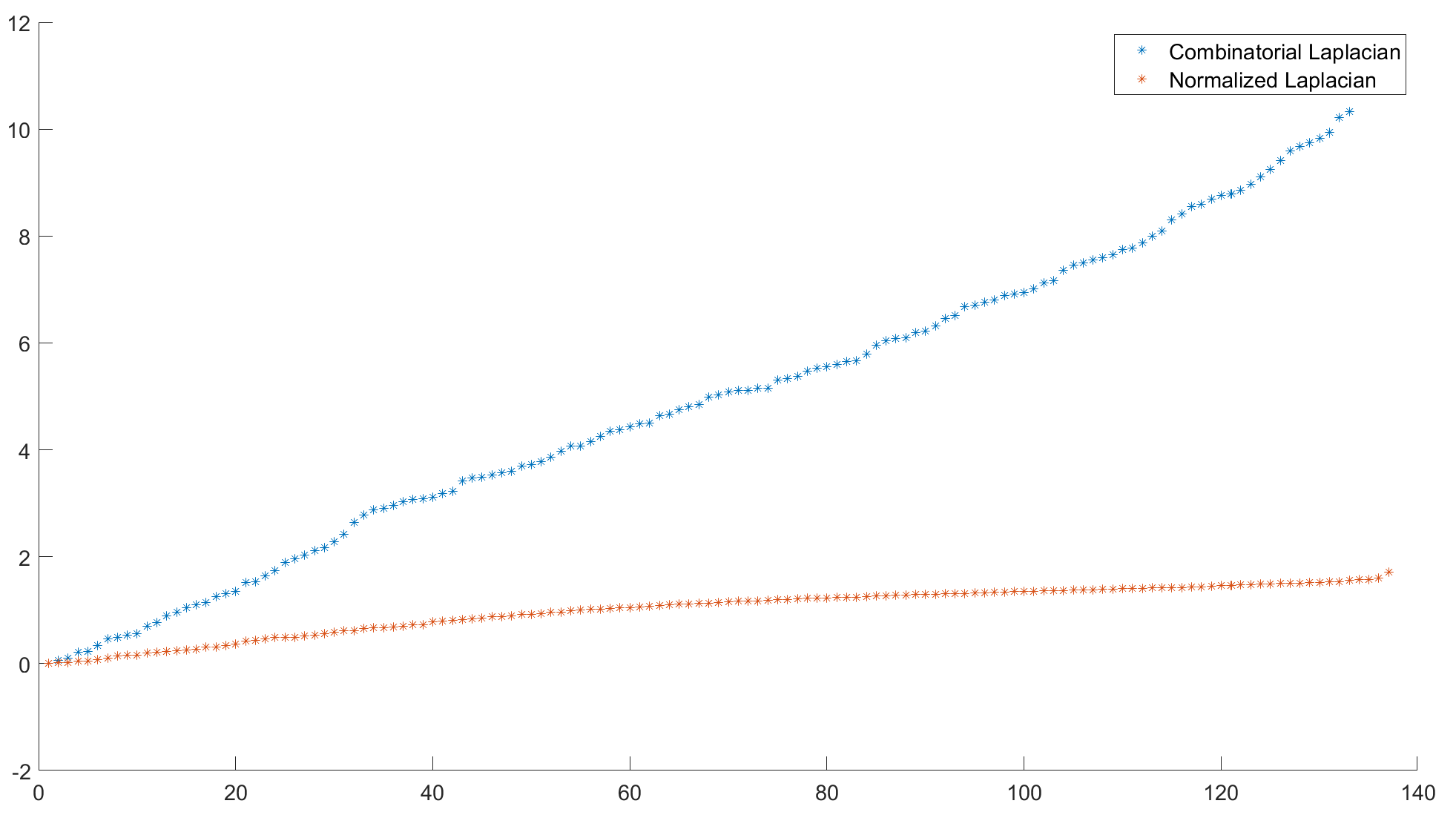}
&\includegraphics[scale=0.16]{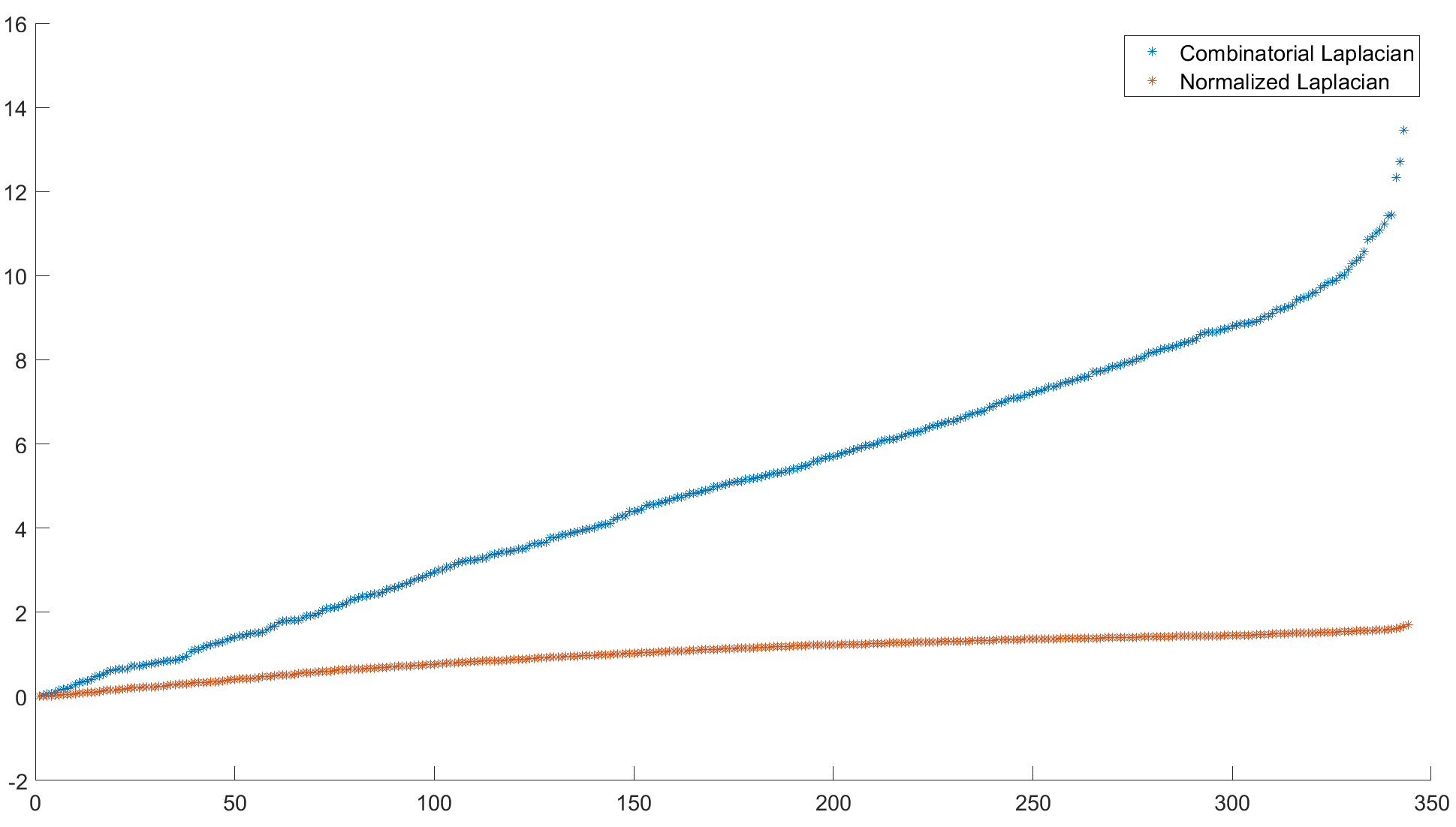}
&\includegraphics[scale=0.16]{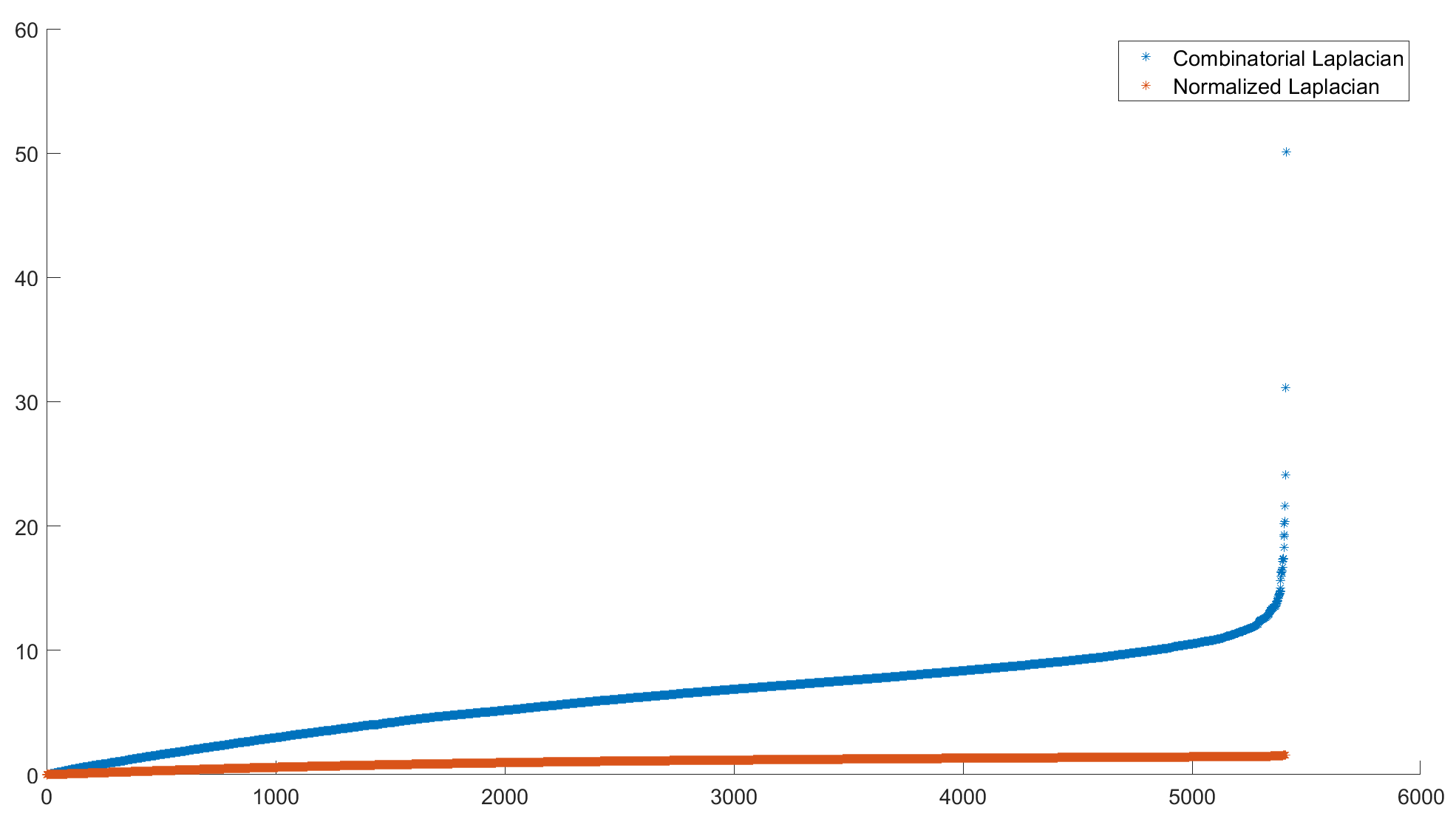}\\
(a) GOA Province
&(b) UK
&(c) NY State
\end{tabular}
\caption{Laplacian eigenvalues of Region Adjacency Graphs. The eigenvalues of the Combinatorial Laplacian are not bounded. The largest eigenvalue (\textit{spectral radius}) increases as the number of nodes increases in the Region Adjacency graph. In contrast, the eigenvalues of the Normalised Laplacian are bounded regardless of the number of nodes in the graph. \label{figure:region_adjacency_graphs_laplacian_eigenvalues}}
\end{figure}

The Combinatorial Laplacian~$\mathbf{L}$ of an undirected graph is symmetric since~$\mathbf{D}$ and~$\mathbf{A}$ are symmetric.~$\mathbf{L}$ is also self-adjoint since~$\mathbf{L}$ is symmetric, in fact, if~$\mathbf{x},\mathbf{y}\in \mathbb{R}^N$ then \mbox{$\langle \mathbf{x},\mathbf{Ly}\rangle=\langle \mathbf{Lx},\mathbf{y}\rangle \Leftrightarrow \mathbf{x}^{\top}\mathbf{Ly}=(\mathbf{Lx})^{\top}\mathbf{y}=\mathbf{x}^{\top}\mathbf{L}^{\top}\mathbf{y}$}. Additionally, since~$\mathbf{L}$ is positive semidefinite, then the Normalised Laplacian~$\hat{\mathbf{L}}$ of an undirected graph is also positive semidefinite because it can be written as \mbox{$\hat{\mathbf{L}}=\mathbf{D}^{-1/2}\mathbf{LD}^{-1/2}$}. Rewriting a Laplacian matrix of a directed graph~$\mathcal{G}=(V,E,\mathbf{A})$ as~$\tilde{\mathbf{D}}-\tilde{\mathbf{A}}$, namely, a diagonal matrix~$\tilde{\mathbf{D}}$ whose~$i$-th element equals the sum of the~$i$-th row of a non-negative matrix~$\tilde {\mathbf{A}}$, then the positive semidefinite property follows from considering the Laplacian as the one corresponding to an undirected graph. For instance, for the Combinatorial Directed Laplacian~$\mathbf{L}_G$ we can set \mbox{$\tilde{\mathbf{A}}=(\mathbf{A}+\mathbf{A}^{\top})/2$} and \mbox{$\tilde{\mathbf{D}}=(\mathbf{D}_{out}+\mathbf{D}_{in})/2$}, which satisfies that the sum of the~$i$-th row equals~$\tilde{D}_{ii}$. Similarly, for the Combinatorial Symmetrised Laplacian~$\mathcal{L}_G$ we have that \mbox{$\tilde{\mathbf{A}}=(\mathbf{\Phi} \mathbf{P}+\mathbf{P}^{\top}\mathbf{\Phi})/2$} and \mbox{$\tilde{\mathbf{D}}=\mathbf{\Phi}$}. The sum of the~$i$-th row of~$\tilde{\mathbf{A}}$ equals~$\Phi_{ii}=\phi_i$, in fact,

\begin{equation*}
    2\sum_{j=1}^N\tilde{A}_{ij}=\sum_{j=1}^N(\Phi \mathbf{P})_{ij}+\sum_{j=1}^N(\mathbf{P}^{\top}\Phi)_{ij}=\sum_{j=1}^N\phi_iP_{ij}+\sum_{j=1}^NP_{ji}\phi_j=2\phi_i,
\end{equation*}

since~$\mathbf{P}$ is row stochastic, and by definition the Perron vector~$\mathbf{\phi}$ satisfies the relation~$\mathbf{\phi}^{\top}=\mathbf{\phi}^{\top}\mathbf{P}$. The positive semidefinite property for the Symmetrised Laplacian follows from the relation \mbox{$\mathcal{L}=\mathbf{\Phi}^{-1/2}\mathcal{L}_G\mathbf{\Phi}^{-1/2}$}. The graph Laplacians for directed graphs, except for the Diplacian, are generally symmetric. Consequently, all their eigenvalues are real. Also, they are bounded for the Symmetrised and the Combinatorial Symmetrised Laplacians (Fig.~\ref{figure:region_adjacency_graphs_laplacian_eigenvalues}).

The Diplacian~$\mathbf{\Gamma}$ of a directed graph is symmetric if and only if its adjacency matrix is symmetric. Indeed, we have~$\mathbf{\Gamma}=\mathbf{I}-\mathbf{\Phi}^{1/2}\mathbf{P\Phi}^{-1/2}$. Assuming~$\mathbf{\Gamma}=\mathbf{\Gamma}^{\top}$, then \mbox{$\mathbf{\Phi}^{1/2}\mathbf{P\Phi}^{-1/2}=\mathbf{\Phi}^{-1/2}\mathbf{P}^{\top}\mathbf{\Phi}^{1/2}$}. Recalling that~$\mathbf{P}=\mathbf{D}^{-1}\mathbf{A}$,~$\mathbf{D}$ and~$\mathbf{\Phi}$ are diagonal matrices, using their multiplication commutativity, and multiplying by~$\mathbf{\Phi}^{1/2}$ from the left and from the right, we obtain \mbox{$\mathbf{\Phi} \mathbf{D}^{-1}\mathbf{A}=\mathbf{A}^{\top}\mathbf{D}^{-1}\mathbf{\Phi}=\mathbf{\Phi} \mathbf{D}^{-1}\mathbf{A}^{\top}$}, Since~$\mathbf{D}^{-1}$ and~$\mathbf{\Phi}$ have an inverse, we get~$\mathbf{A}=\mathbf{A}^{\top}$ as a necessary condition for the symmetry of the Diplacian. Conversely, if~$\mathbf{A}=\mathbf{A}^{\top}$ then~$\mathbf{\Gamma}$ is symmetric. If~$\mathbf{A}$ is symmetric, then the Diplacian reduces to~$\mathbf{I}-\mathbf{P}=\mathbf{I}-\mathbf{D}^{-1/2}\mathbf{AD}^{-1/2}=\hat{\mathbf{L}}$, i.e. the Normalised Laplacian.

\subsection{Graph Laplacians and traffic models}
The Graph Laplacian can be used to estimate the population of the sub-regions in a given area for which mobility or traffic flows are known. In~\citep{metro_scalzo} it is proposed a way to estimate the population distribution~$\phi$ surrounding the stations of the London underground network, by using Fick's law of diffusion~\mbox{$\mathbf{q}=-k\nabla\mathbf{\phi}$} which states that the flux~$\mathbf{q}$ flows from regions of high concentration to regions of low concentration, with a magnitude proportional to the concentration gradient~\mbox{$\nabla \mathbf{\phi}$}, and coefficient of diffusivity~$k$. Because of the graph structure and the discrete nature of the stations, an algebraic manipulation of the Fick's law gives the relation~\mbox{$\mathbf{q}=-k\mathbf{L\phi}$}, where~$\mathbf{L}$ is the combinatorial Laplacian of the network. The flux~$\mathbf{q}$ was estimated using the average daily flow of passengers obtained from Transport for London data. Using the pseudo-inverse matrix~$\mathbf{L}^+$, the estimation of the population surrounding each station is obtained through~$\hat{\mathbf{\phi}}=-(1/k)\mathbf{L}^+\mathbf{q}$. 

\section{Conclusions, data sets and future work\label{sec:FUTURE-WORK}}
The data for defining the traffic features include sensors, GPS, rail-hailing, and transaction data sets. Sensors can be implemented on roads and highways and collect traffic measurements, such as speed. GPS trajectories are generated by taxis or rented vehicles during a period. Rail-haling records include some transport services demand, such as rented bicycles or mobility as a service option. Transaction data sets are generated by automatic systems, such as ticket machines in metro stations, which count the number of departing or arriving passengers.

Graph-based structures allow the analysis of the interplay between the different elements of the system, as well as the identification of the most relevant or vulnerable nodes within the whole network. This type of method is even applicable for some grid-based structures in traffic and mobility modelling as long as Origin-Destination matrices are available, as in the case of the Deep Gravity model. Furthermore, node centrality metrics are an instrument of structure analysis of the network even if there are no available mobility flows, as in the case of the Region Adjacency graphs. Similarly, the Laplacian matrix can provide additional reasoning about external information, e.g., the distribution of the population surrounding the nodes of an urban network. For connected and strongly connected graphs, the existence and uniqueness of the Perron vector associated with the transition probability matrix enables the definition of several Laplacian matrices and constructs a circulation function in the graph, giving more quantitative information for analysis and visualisation.

In future work, we plan to address the monitoring of mobility data to reconstruct traffic flows in urban, regional and inter-regional contexts through the analysis of origin-destination data, i.e., of mobility trajectories, considering heterogeneous, partial and uncertain data, and integrating them with meteorological and pollution data. In particular, it will be possible to analyse the analysis of short-term mobility patterns to classify the mobility habits of users, e.g., assessing whether a particular route/access is regular or sporadic. Furthermore, an implementation of a graph-based Deep Gravity mobility model with a strongly connected structure is desired to implement the circulations induced by the Perron vector of the transition probability matrix, as well as the exploration of alternatives for the circulation function when the graph is not connected or strongly connected, for instance, based in node centrality metrics.

\textbf{Data availability}. The Python code, the Region Adjacency data and the predicted flows using the Deep Gravity mobility model~\citep{deepgravity_pappalardo} for New York State are freely available at \url{github.com/scikit-mobility/DeepGravity}. The Region Adjacency data for the United Kingdom and Genova Province are freely available at \url{https://census.ukdataservice.ac.uk/use-data/guides/boundary-data}, and \url{https://smart.comune.genova.it/opendata}, respectively.

\textbf{Declarations of interest}. None.

\textbf{Funding}. Rafael Martínez Márquez has been supported by a REACT-EU PhD fellow.

\bibliographystyle{apalike}
\bibliography{bib}

\end{document}